\documentclass[aps,prb,twocolumn,superscriptaddress,showpacs]{revtex4-2}
\usepackage{amsmath,amssymb,amsfonts,amsthm}
\usepackage{graphicx}
\usepackage{bm}
\usepackage{bbm}
\usepackage{color}
\usepackage{dcolumn}   
\usepackage{epstopdf}
\usepackage[caption=false]{subfig}
\usepackage{xr}
\usepackage{color}
\usepackage[colorlinks=true,linkcolor=blue,urlcolor=blue,citecolor=blue]{hyperref}

\definecolor{darkblue}{rgb}{0, 0, 0.8}

\begin{document}

 \newcommand{\breite}{1.0} 

\newcommand{\beq}{\begin{equation}}
\newcommand{\eeq}{\end{equation}}

\newcommand{\bea}{\begin{eqnarray}}
\newcommand{\eea}{\end{eqnarray}}
\newcommand{\lt}{<}
\newcommand{\gt}{>} 

\newcommand{\Reals}{\mathbb{R}}     
\newcommand{\Com}{\mathbb{C}}       
\newcommand{\Nat}{\mathbb{N}}       

\newcommand{\id}{\mathbboldsymbol{1}}    

\newcommand{\Real}{\mathop{\mathrm{Re}}}
\newcommand{\Imag}{\mathop{\mathrm{Im}}}

\def\O{\mbox{$\mathcal{O}$}}   
\def\B{\mbox{$\mathcal{B}$}}   
\def\C{\mbox{$\mathcal{C}$}}   
\def\F{\mathcal{F}}			
\def\sgn{\text{sgn}}

\newcommand{\deo}{\ensuremath{\Delta_0}}
\newcommand{\Po}{\ensuremath{\ket{\Psi_o}}}
\newcommand{\Pe}{\ensuremath{\ket{\Psi_e}}}
\newcommand{\dea}{\ensuremath{\Delta}}
\newcommand{\aj}{\ensuremath{a_j}}
\newcommand{\ajd}{\ensuremath{a^{\dagger}_{j}}}
\newcommand{\sx}{\ensuremath{\sigma_x}}
\newcommand{\sz}{\ensuremath{\sigma_z}}
\newcommand{\spl}{\ensuremath{\sigma_{+}}}
\newcommand{\smi}{\ensuremath{\sigma_{-}}}
\newcommand{\alk}{\ensuremath{\alpha_{k}}}
\newcommand{\bk}{\ensuremath{\beta_{k}}}
\newcommand{\om}{\ensuremath{\omega}}
\newcommand{\dw}{\ensuremath{\Delta_0}}
\newcommand{\wbp}{\ensuremath{\omega_0}}
\newcommand{\dv}{\ensuremath{\Delta_0}}
\newcommand{\vbp}{\ensuremath{\nu_0}}
\newcommand{\vplus}{\ensuremath{\nu_{+}}}
\newcommand{\vminus}{\ensuremath{\nu_{-}}}
\newcommand{\wplus}{\ensuremath{\omega_{+}}}
\newcommand{\wminus}{\ensuremath{\omega_{-}}}
\newcommand{\uv}[1]{\ensuremath{\mathbf{\hat{#1}}}} 
\newcommand{\dg}{\ensuremath{\dagger}}

\newcommand{\lr}[1]{\left( #1 \right)}
\newcommand{\lrs}[1]{\left( #1 \right)^2}
\newcommand{\lrb}[1]{\left< #1\right>}

\newcommand{\ket}[1]{\left| #1 \right>} 
\newcommand{\bra}[1]{\left< #1 \right|} 
\newcommand{\braket}[2]{\left< #1 \vphantom{#2} \right|
 \left. #2 \vphantom{#1} \right>} 
\newcommand{\matrixel}[3]{\left< #1 \vphantom{#2#3} \right|
 #2 \left| #3 \vphantom{#1#2} \right>} 

\newcommand{\pdd}[2]{\frac{\partial^2 #1}{\partial #2^2}} 
\newcommand{\pdc}[3]{\left( \frac{\partial #1}{\partial #2}
 \right)_{#3}} 
 \renewcommand{\d}[2]{\frac{d #1}{d #2}} 
\newcommand{\dd}[2]{\frac{d^2 #1}{d #2^2}} 
\newcommand{\pd}[2]{\frac{\partial #1}{\partial #2}} 
\newcommand{\grad}[1]{{\nabla} {#1}} 
\let\divsymb=\div 
\renewcommand{\div}[1]{{\nabla} \cdot \boldsymbol{#1}} 
\newcommand{\curl}[1]{{\nabla} \times \boldsymbol{#1}} 
\newcommand{\laplace}[1]{\nabla^2 \boldsymbol{#1}}
\newcommand{\vs}[1]{\boldsymbol{#1}}
\newcommand{\abs}[1]{\left| #1 \right|} 
\newcommand{\avg}[1]{\left< #1 \right>} 

\let\baraccent=\= 

\title{Signatures of Majorana Zero-Modes in an isolated one-dimensional superconductor
}

\author{Rohith Sajith}
\affiliation{Department of Physics, University of California, Berkeley, CA, USA}
\email{rohithsajith@berkeley.edu}
\author{Kartiek Agarwal}
\affiliation{Department of Physics, McGill University, CA}
\author{Ivar Martin}
\affiliation{Material Science Division, Argonne National Laboratory, Argonne, IL 08540, USA}
\email{ivar@anl.gov}

\date{\today}
\begin{abstract}
We examine properties of the mean-field wave function of the one-dimensional Kitaev model supporting Majorana Zero Modes (MZMs) \emph{when restricted} to a fixed number of particles. Such wave functions can in fact be realized as exact ground states of interacting number-conserving Hamiltonians and amount to a more realistic description of the finite isolated superconductors. Akin to their mean-field parent, the fixed-number wave functions encode a single electron spectral function at zero energy that decays exponentially away from the edges, with a localization length that agrees with the mean-field value. 
Based purely on the structure of the number-projected ground states, we construct the fixed particle number generalization of the MZM operators. They can be used to compute the edge tunneling conductance; however, notably the value of the zero-bias conductance remains the same as in the mean-field case, quantized to $2e^2/h$.
We also compute the topological entanglement entropy for the number-projected wave functions and find that it contains a `robust' log(2) component as well as a logarithmic correction to the mean field result, which depends on the precise partitioning used to compute it. The presence of the logarithmic term in the entanglement entropy indicates the absence of a spectral gap above the ground state; as one introduces fluctuations in the number of particles, the correction vanishes smoothly.

\end{abstract}
\maketitle

\section{Introduction} 
Majorana fermions are the real solutions of the Dirac equation that act as their own antiparticles. Remarkably, in the condensed matter setting, Majorana fermions emerge as natural quasiparticles in magnetic \cite{2chK, kitaev2003fault, kitaev2006anyons} 
and superconducting systems that exhibit topological order in one~\cite{kitaev2001unpaired} and two dimensions~\cite{read2000paired, ivanov2001non}.
While the usual Dirac, or complex, fermions can always be decomposed into pairs of Majorana fermions, it is only in certain cases, when a system has topological order, that one can realize spatially {\em unpaired} Majorana fermions as quasiparticles. These states commute with the Hamiltonian and thus cost zero energy; they encode the topologically protected ground state degeneracy of the system and are referred to as Majorana zero modes (MZMs). 
{Unlike complex fermions or Abelian anyons whose exchange only results in a phase transformation of the wave function, MZMs exhibit non-Abelian exchange statistics, whereby their exchange results in a unitary transformation on the multi-dimensional ground state manifold.}
This makes MZMs a valuable component of putative quantum computers operating on a quantum register of qubits encoded in the ground state degeneracy of a topological many-body system~\cite{nayak2008non}.  

While MZMs can be built into certain interacting spin models {\em exactly}, {they are realized in superconductors as zero-energy self-conjugate Bogoliubov quasiparticles $\gamma$ \cite{ivanov2001non, kitaev2001unpaired} satisfying $\gamma ^2 = 1$ within Bardeen-Cooper-Shrieffer (BCS) mean field theory}. At the mean-field level, superconductors have a well-defined phase that is conjugate to the number of electrons; thus, the ground state has a fluctuating number of electrons. If we consider an electrically isolated piece of superconductor, such fluctuations are clearly impossible. Therefore, strictly speaking, the mean-field description cannot be correct in a finite system, and the survival of MZMs in this setting becomes a non-trivial problem. 

While this could be a matter of concern even in large superconductors, it is particularly critical in thin-wire superconductors where phase fluctuations are further enhanced due to reduced spatial dimensionality and system-size effects. 
Given that this is precisely the setting of some topological quantum computing schemes based on manipulation of MZMs \cite{lin2018towards, lutchyn2018majorana}, it is important to carefully examine the consequences of going beyond the BCS mean-field limit.

To address the concerns about possible artifacts of the BCS approximation on Majoranas, we examine the presence of MZMs in a one-dimensional superconducting chain by shifting focus from the Hamiltonian to the structure of the many-body ground state. At the mean-field level, the Kitaev p-wave superconducting chain has two MZMs in its topological phase, one at each edge of the superconductor. Instead of examining the full mean-field Kitaev ground state with a fixed phase and a fluctuating number of electrons, we consider the states obtained when $\ket{\Psi_K}$ is projected onto a fixed number $N$ of electrons, $\ket{N}$.  

It may appear that the number projection procedure on the BCS wave function is rather arbitrary and not guaranteed to give a good approximation of the many-body wave function in a superconductor with a fixed number of particles. One can show, however, that the number projection procedure gives the same result as a variational calculation of a fixed-number wave function using a number-conserving interacting Hamiltonian \cite{leggett2006quantum}.
In the case of Kitaev wave function, in fact, it is possible to explicitly construct a number-conserving Hamiltonian for which $\ket{\Psi_K}$ is the {\em exact} ground state~\cite{Iemini15, WangNumberConserving}. The Hamiltonian is physically meaningful, with only short-range hopping and interactions. Since the Hamiltonian does not mix different number sectors, the fact that $\ket{\Psi_K}$ is the ground state, automatically implies that that all projections $\ket{N}$ are the ground states as well. This gives us an additional reason to study the properties of $\ket{N}$ in detail. 

{We generally find that the number-projected version of the Kitaev wave function indeed retains some key features typically associated with MZMs. Namely, the single-electron spectral function has a zero-frequency peak near the edges of the wire, in direct analogy to the mean-field MZMs. We are also able to construct a proper generalization of Majorana operators for the fixed number case, which induces exact transitions between ground states that differ by one in the number of electrons, $|N\rangle \leftrightarrow |N+1\rangle$.
Similar to the standard mean-field Majorana operator, this operator (superficially) appears local. However, in reality, it encodes non-local correlations via a Cooper pair operator that it explicitly contains. The Cooper pair $P^{\dagger}$ induces a transition from the state $\ket{N}$ to the state $\ket{N + 2}.$ The form of the Majorana operators happens to match the conjecture made recently in Ref.~\cite{lin2018towards}.}

Focusing exclusively on the many-body ground-state wave function allows us to make Hamiltonian-independent statements. In this way, our approach is complementary to the exact solutions of bulk models of topological superconductors available in some cases \cite{Ortiz14, Iemini15, Lang2015}, bosonization analysis \cite{Sau2011, Cheng11, Lukasz2011, Ruhman2015}, and DMRG \cite{Kraus2013, Keselman2015}. However, it also makes it impossible to access some important quantities such as the gap between the ground state(s) and the excited states. {One can partially address this issue by studying the entanglement properties of the wave function. For the projected Kitaev wave function $\ket{N}$, we find that the topological entanglement entropy exhibits a robust $\log(2)$ value, identical to that observed for the mean-field wave function in the topological phase. However, it additionally contains a logarithmic correction that is dependent on the precise geometry of the partitions used to compute the topological entanglement entropy. These results suggest that such a wave function can only appear as the ground state of a {\em gapless} Hamiltonian~\cite{hastings2007area}. Although this does not completely preclude the presence of topologically protected zero modes~\cite{verrsengaplesstopo}, it may make the dynamical manipulation of putative MZMs challenging. We leave the numerical study of braiding and measurement-based computing with MZMs for future work.}

\section{fixed number wave function}
\subsection{Mean field model and its ground states.}

Our main object of study is the number-projected ground state wave function of Kitaev's model for the mean-field $p$-wave superconductor \cite{kitaev2001unpaired}. To start, let us summarize the main points about the mean-field model. Its Hamiltonian is
\begin{equation}
H_{MF} = -\sum_{j = 1}^{L - 1} \{ t a_{j}^{\dg}a_{j + 1} + \mu a_{j}^{\dg}a_{j} - \Delta a_{j}a_{j + 1} + h.c. \} \label{eq:HMF}
\end{equation}
where $a_{j},$ $a^{\dg}_{j},$ and $n_{j}$ are fermionic annihilation, creation, and density operators for the $j$-th site, $\Delta$ is the superconducting gap, $t$ is a hopping amplitude, $\mu$ is the chemical potential, and $L$ is the chain length.

The topological phase persists as long as $|\mu|\lt 2t$ and is characterized by the appearance of Majorana modes near the system's edges when placed on open boundary conditions. 
For $\Delta = t$,  and $\mu = 0$, they are perfectly isolated on the first and the last sites of the chain, and can be expressed in terms of the physical electrons as $\gamma_1 = a_1 + a_1^\dagger$  and $\gamma_2= -i (a_L - a_L^\dagger)$.
These operators, as well as the corresponding complex fermion $f =( \gamma_1 + i\gamma_2)/2,$ have a trivial Heisenberg evolution. This implies that $f$ is a zero-energy fermion mode. Its occupation number $n_f = f^\dagger f = \{0,1\}$ can be used to label the single wire ground states. 
Explicitly, the two ground states of the Hamiltonian at the special $t = \Delta$ point of (\ref{eq:HMF}) are \cite{alicea2011non}
\beq
\ket{\Psi_{e,o}} = \frac 1 {2^{\frac{L-1} 2}}\sum_{n_1+\ldots+n_L = e,o} {\lr {a_1^\dag}}^{n_1}{\lr {a_2^\dag}}^{n_2}\ldots{\lr {a_L^\dag}}^{n_L}\ket 0. \label{eq:Psi}
\eeq
Note that the sum goes over any combination $\{n_j\}$ such that the total number of electrons is either odd or even, depending on the sector.
Indeed,
$
H_{MF}\ket{\Psi_{e,o}} = -({L-1})\ket{\Psi_{e,o}};
$
that is, odd and even states have the same energy.  Moreover, $f\ket{\Psi_e} = 0$ and $f^\dag\ket{\Psi_e} = \ket{\Psi_o}$ -- the odd and even states are the eigenstates of $n_f$ with the eigenvalues 1 and 0, respectively.


\subsection{Wave function projected to fixed number of particles}
The mean-field BCS wave function is a superposition of states with different numbers of electrons and thus cannot be literally correct for an isolated system.  An alternative variational treatment in a fixed-number sector, however,
yields the same  BCS equations for the transition temperature and the gap equation \cite{leggett2006quantum}. As one could anticipate, the fixed-number {\em generalized} BCS wave function is nothing but the mean-field BCS wave function, projected onto a fixed number of particles.

Applied to the Kitaev chain, this procedure yields
\beq
\ket{{N}} = \frac{1}{\sqrt{ L\choose N}}\sum_{n_1+\ldots+n_L = N} {\lr {a_1^\dag}}^{n_1}{\lr {a_2^\dag}}^{n_2}\ldots{\lr {a_L^\dag}}^{n_L}\ket 0, \label{eq:PsiN}
\eeq
\textcolor{black}{Note the a change in normalization factor since ${L\choose N} = \frac{L!}{N!(L-N)!}$ is  the number of  configurations with $N$ electrons.} Even though the wave functions $\ket{{N}}$ are obtained  from the mean-field wave functions $\ket{\psi_{e, o}}$, we are interested in identifying Majorana-like features contained in $\ket{{N}}$, irrespective of their mean-field origin. 

The eigenstates $\ket{{N}}$ could originate from a variety of Hamiltonians. A particularly nice example is a fully conserving Hamiltonian of $N$ spinless fermions hopping on an $L$ site one-dimensional wire with open boundary conditions constructed in Refs. \cite{Iemini15,WangNumberConserving}: 
\begin{equation}
    H = -J \sum_{i = 1}^{L - 1} \{ a^{\dg}_{j}a_{j + 1} + a^{\dg}_{j + 1}a_{j} - n_{j} - n_{j + 1} + 2 n_{j} n_{j + 1} \}.
\label{eq:HHazzard}
\end{equation}
 Via a Jordan-Wigner Transformation, this Hamiltonian can be written as a spin-$1/2$ ferromagnetic Heisenberg chain. The ground states are fully polarized states with a total spin of $L/2$. It has the degeneracy $L+1$ due to the arbitrary orientation of the total moment (number of distinct projections of the total moment on any given axis). The energy gap between the degenerate ground states and the first excited state corresponds to single magnon excitations and hence scales as $L^{-2}$.
 The $(L + 1)$-fold ground state degeneracy of the Heisenberg model corresponds to the ground state degeneracy across $L + 1$ possible number sectors in the fermion picture.

 While not important for most of the present work, having a number conserving Hamiltonians $H$ will be needed when we study a junction-type braiding protocol in future work; here we focus exclusively on the properties of $\ket{{N}}.$ 
 
{A final note. In this study, we restrict ourselves to the parent mean-field wave function $\ket{\Psi_K}$ constructed for $\mu = 0$ due to its simplicity. Nevertheless, we will use it to access $\ket{N}$ states with $N$ values that correspond to the filling fraction $p$ other than half-filling. Despite this simplification, the localization length of MZMs predicted by this wave function agrees remarkably well with the mean-field result even away from half-filling for a large range of $p$.}

\section{Properties of projected wave function}
The mean-field solution of the Kitaev Hamiltonian has a Bogoliubov quasiparticle at zero energy (i.e., at chemical potential) with probability amplitude concentrated near both ends of the chain. This quasiparticle leads to a zero-energy peak in the density of states while tunneling into the edge sites, but not into the bulk. 
Each of the edge modes in the mean-field treatment is associated with a Majorana zero mode. 

In this section, we will see how these features manifest in the number-projected wave function. We find that the spectral function retains the zero-energy peaks near the edges and that it is possible to construct operators analogous to the Majorana operators that induce transitions between ground states with $N$ and $N+1$ particles, perfectly in the limit of $L, N \to \infty$. In the process, we also construct a Cooper pair operator, which switches between states $N$ and $N+2$.


\subsection{Edge mode and spectral function}\label{sec:spec}

The hallmark of the edge Majorana modes in Kitaev wire is the appearance of the peak in spectral function at zero energy near the wire edges.  In the mean-field treatment, this originates from the self-conjugate Bogoliubov quasiparticles at the edge. In the many-body setting such quasiparticles a priori may not exist, but the spectral function can be computed for any number-projected ground states $\ket{N}$.  It is defined as
\beq
A_i(\omega) = \sum_n |\bra {\psi_n} a_i^\dagger \ket N|^2 \delta(\omega - E_n + E_N),
\eeq
where the sum goes over all states $\ket {\psi_n}$ connected to $\ket N$ by a single electron creation operator.
Let us examine matrix elements for the transitions between ground states; thus, we may set $\ket {\psi_n} = \ket{N+1}$.
As an example, suppose we try to add an electron to site 1.  The result is 
\beq
a_1^\dag\ket{N} = \frac{1}{\sqrt{ L\choose N}}\sum_{n_2+...+n_L = N} a_1^\dag {\lr {a_2^\dag}}^{n_2}{\lr {a_3^\dag}}^{n_3}...{\lr {a_L^\dag}}^{n_L}\ket 0.\label{eq:PsiN}
\eeq
{There are $ {L-1}\choose N$ terms in this sum. The overlap with the number-projected state with $N+1$ electrons is therefore
\beq
\bra{{N+1}}a_1^\dag\ket{N} = \frac{{{L-1}\choose N}}{\sqrt{{{L}\choose N} {{L}\choose {N+1}}}} \to \sqrt{p(1-p)}, \label{eq:a1}
\eeq
the latter valid in the limit of large $L$ and $N$, and finite $\frac{N}{L}\equiv p.$  By symmetry, the same result holds for the matrix element of $a^\dag_L$. 

We may also compute the amplitude to insert an electron at an arbitrary site $j,$  $\bra{{N + 1}} a_{j}^{\dg} \ket{{N}}.$ Due to the anticommutation of fermion operators, we may express this as a sum over the number of fermions $k$ that are present at the sites $1\le k \le j-1$ 

\beq
\bra{{N + 1}} a_{j}^{\dg} \ket{{N}} = \sum_{k = 0}^{j-1} (-1)^{k}  \frac{{{j-1}\choose k} {{L - j}\choose {N - k}}}{\sqrt{{{L}\choose N} {{L}\choose {N+1}}}}
\eeq
In the limit of large system size and $j \ll L$, the above expression  simplifies as 
\beq
\begin{split}
\bra{{N + 1}} a^{\dg}_{j} \ket{{N}} &= \frac{{{L-1}\choose N}}{\sqrt{{{L}\choose N} {{L}\choose {N+1}}}} \\
&\times\sum_{k = 0}^{j - 1} {j-1 \choose k}(-1)^{k} p^{k} (1-p)^{j - 1 - k} \\
&= \sqrt{p(1-p)} (1-2p)^{j-1}\label{eq:Mj}
\end{split}
\eeq
The numerator in the prefactor, just as before, is the number of states where one of $N+1$ electrons is fixed in the lattice; the terms under the sum correspond to the probabilities to have $k$ electrons in the first $j-1$ sites, with the sign determined by the parity of $k$. Note that below half-filling ($p \lt 0.5$) the matrix element has a positive sign, while above half-filling, it is oscillatory. Exactly at half-filling, it is only possible to create an electron on the first site without exciting outside the ground-state manifold. Ignoring the sign changes, the general expression of the inverse decay length is $\xi_p^{-1} = \ln |1 - 2p| $. 
In Fig.~\ref{fig:decay}, we compare the exact combinatorial evaluation of the matrix element and the limiting result of Eq. (\ref{eq:Mj}) for a system of total length $L = 500$.  The agreement is very good near the edges of the wire. 

We can further compare the localization lengths for the number-projected wave functions with those of the mean-field ground states of the Kitaev Hamiltonian. 
At the special point with $t = \Delta$, the localization length of Majorana edge modes in the topological phase is determined by  $\xi^{-1} = \ln \frac{|\mu|}{2t}$~\cite{kitaev2001unpaired}. The chemical potential $\mu$ is  related to average electron density; by diagonalizing the mean-field Hamiltonian, we find
\beq
p (\mu) = \int_0^{2\pi}\frac{dk}{4\pi} \left[1 + \frac{t\cos k + \mu}{\sqrt{t^2 + \mu^2 + 2t\mu\cos k}}\right].
\label{eq:pmu}
\eeq

In a finite-length chain, this value of filling $p$ is only defined approximately due to the uncertainty in the number of particles that scales as $\sqrt{N}$ in the mean-field ground state. The fluctuation of fillings $\Delta p \sim L^{-0.5}$ is also finite at intermediate fillings.  Thus, for large but finite system sizes, the mean-field wave function at $p\ne 0.5$ can be quite different from the one $p = 0.5$ (chemical potential $\mu = 0$), which we use as the parent wave function for our number-projected ground states. 
Despite this difference, a direct comparison of the mean-field Majorana localization length as a function of density and the decay length of the single electron matrix element 
between the number-projected ground states 
[Eq. (\ref{eq:Mj})] shows an excellent match in a finite range of fillings near $p= 0.5$; see Fig. {\ref{fig:NMF}}.

\begin{figure}
\includegraphics[width=3.3in]{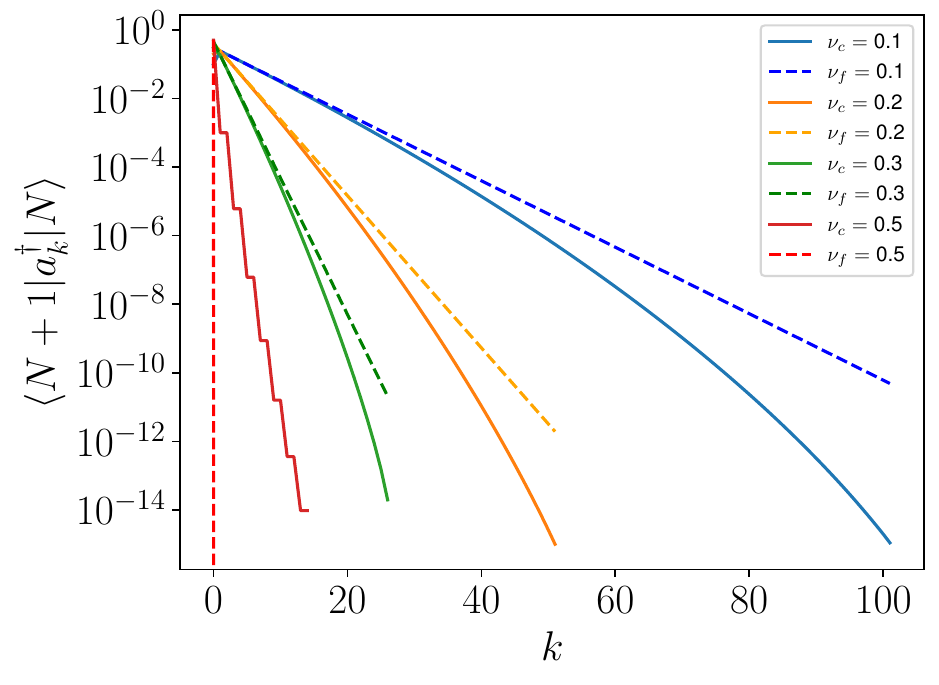}
\caption{Matrix element $\bra{\Psi_{N + 1}} a^{\dg}_{k} \ket{\Psi_{N}}$ for different fillings computed using the exact combinatorial expression for $L = 500.$ The exponentially decaying amplitude for a single-electron matrix element between $\ket{\Psi_{N}}$ and $\ket{\Psi_{N + 1}},$ is strongly reminiscent of the  Majorana edge mode in the mean-field model. We further fit the combinatorial result with the prediction from \eqref{eq:Mj}  (dashed lines) and find good agreement near the edge of the wire.}\label{fig:decay}

\end{figure}

\begin{figure}
\includegraphics[width=3.3in]{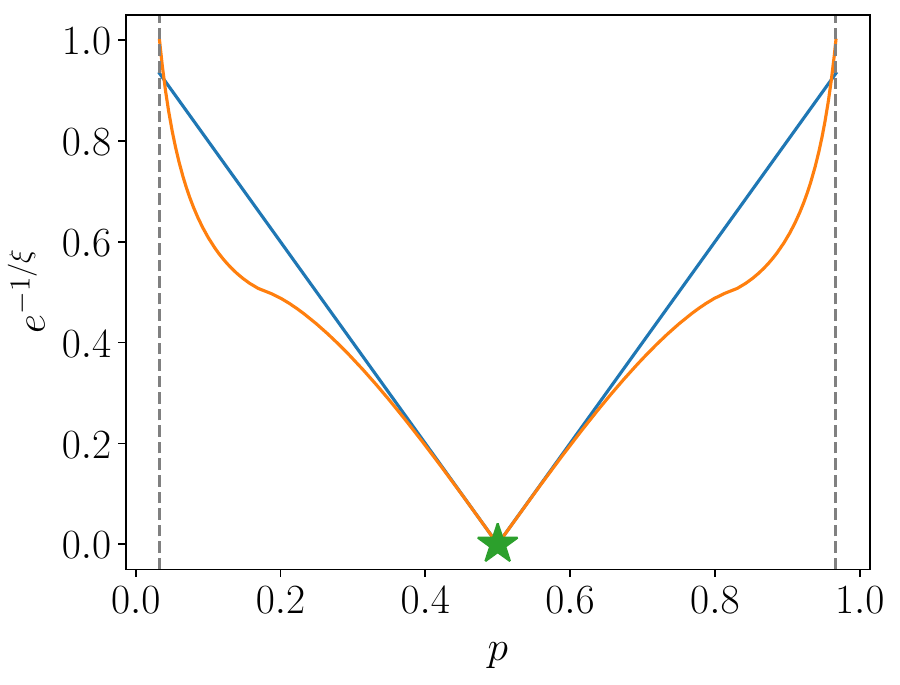}
\caption{Comparison of localization lengths $\xi$ of Majorana zero-modes for the mean-field Kitaev model (orange) and number-projected model (blue) as a function of filling fraction $p$ at $t = \Delta = 1.$ Near half-filling, we see strong agreement between the mean-field and number-projected predictions. The gray dashed lines indicate the boundaries of the topological phase for the Kitaev model.}\label{fig:NMF}
%
\end{figure}

Away from $p = 0.5$  and towards the boundaries of the mean-field topological phase ($\mu = \pm 2t$), the discrepancy between the strictly linear relation produced by the number-conserving wave functions and the Kitaev model grows. Also, while the Kitaev model only has a topological phase within a fixed range of filling fractions, this number-conserving scheme finds evidence of Majorana edge modes down to arbitrarily low or high filling in the infinite system limit.

Turning back to Eq. (\ref{eq:Mj}), we note that creating an electron at the edge site provides an $\mathcal{O}(1)$ matrix element for transition between ground states with $N$ and $N+1$ electrons, which have opposite parity. In this sense, the single electron operator acts similarly to the intended behavior of Majorana operators subject to the constraint that states must have fixed number of electrons. 

Building upon this observation, we next construct an operator $\Gamma$ that converts between number-projected states perfectly, satisfying $\ket{N+1}= \Gamma^\dagger\ket{N}.$ This is in contrast to $a_1^\dagger$ which only gives a $\mathcal{O}(1)$ matrix element as demonstrated in Eq. \ref{eq:a1}. The key ingredient to construct $\Gamma^\dagger$ is the Cooper pair operator $P^\dagger$, which accomplishes the transition between states $\ket{N+2} = P^\dagger \ket{N}$. 

\subsection{Cooper pair creation operators at fixed N}

We define the Cooper pair creation operator as the operator that transforms $\ket{\Psi_{N}}$ into $\ket{\Psi_{N + 2}}.$ It is simple to see that the ansatz 
\beq
    P^{\dg} = \sum_{i = 1}^{L - 1} a^{\dg}_{i}a^{\dg}_{i + 1}\label{eq:P}
\eeq
accomplishes precisely that for $L,N\to \infty$, since  
\beq
\frac{\bra{{N + 2}} P^{\dg} \ket{{N}}}{\sqrt{\bra{{N}} P P^{\dag} \ket{{N}}}}  \rightarrow 1. \label{eq:Pnorm}
\eeq
That is, the normalized state $P^{\dg} \ket{{N}}$ becomes identical to $ \ket{{N+2}}$. To leading order in $L$, Eq. (\ref{eq:Pnorm}) can be written as ${ \binom{L - 2}{N}}/{\sqrt{\binom{L}{N + 2} \binom{L - 4}{N - 2}}}$, which indeed approaches 1 for any finite $p$ as $L\to \infty$.

Viewing $\ket{{N}}$ as a superposition of bitstrings, it might seem surprising that $P^{\dg}$ creates all bitstrings with $N + 2$ particles since we only add particles on adjacent sites. However, the set of $N+2$ particle states that are reachable from $\ket{N}$ via $P^{\dg}$ are those that have at least 1  pair of adjacent particles somewhere within the system. At fixed filling $ p = {N}/{L}$ and $N, L \to \infty$, all random bit strings have such a local pair somewhere within the system with probability $1$, meaning that $P^{\dg}$ indeed reaches all $N + 2$ particle states in the large system limit.

Interestingly, the Cooper pair operator of Eq. (\ref{eq:P}) is not unique. It is easy to show that
\beq
    P^{\dg}_\ell = \sum_{i = 1}^{L-\ell} a^{\dg}_{i}a^{\dg}_{i + \ell}\label{eq:P}
\eeq
also works. First of all, for $L, N \to \infty$
\beq\bra{N+2} a^{\dg}_{i}a^{\dg}_{i + \ell} \ket{N} = p(1-p)(1-2p)^{\ell -1}.\label{eq:Pj}
\eeq
Note that the decay law of this anomalous correlator (\ref{eq:Pj}) is the same as the matrix element for a single-electron zero-energy transition between number-projected states in Eq. (\ref{eq:Mj}). Only for $p = 0.5$ this matrix element vanishes for $\ell > 1$. Normalizing in the same way as Eq. (\ref{eq:Pnorm}) shows that $P^\dagger_\ell \ket{N}= \sgn(1-2p)^{\ell-1}\ket{N+2}$. 

The redundancy in the definition of the Cooper pair operator thus implies that they all act identically in the ground state manifold of the number-projected states. This  follows from Eq. (\ref{eq:Pnorm}): 
$${\bra{{N}} P\ket{{N+2}}}{\bra{{N + 2}} P^{\dg} \ket{{N}}}={{\bra{{N}} P P^{\dag} \ket{{N}}}}.$$
Inserting the resolution of identity on the right-hand side shows that the operator $P^\dg$ only connects one ground state to another ground state with two additional electrons.

This perfect transformation also implies that a linear superposition with arbitrary amplitudes $\alpha_\ell$,  $\sum_\ell \alpha_\ell P_\ell$ is also a legitimate Cooper pair operator~(see Ref.~\cite{WangNumberConserving}, where $\alpha_l = \text{const.}$).   

In what follows, we will assume that the Cooper pair operators are normalized, such that $\ket{{N + 2}} = P^{\dg} \ket{{N}}.$

\subsection{Majorana Operators at fixed N}
We are now ready to define the Majorana operators $\Gamma^\dagger$ for the number-projected case as an operator that induces a perfect transition between ground states with $N$ and $N+1$  electrons:
\beq
\frac{\bra{N+1} \Gamma^\dg \ket{N}}{\sqrt{\bra{N} \Gamma\Gamma^\dg \ket{N}}}\to 1. 
\eeq
Motivated by the mean-field analogy and the result in Eq. (\ref{eq:Mj}), we look for operators of the form $\Gamma_L^\dg = \sum_{j = 1}^L \beta^{j-1} (a_j^\dg + a_j P^\dg)$ at the left edge of the wire, and analogously, $\Gamma_R^\dg = i \sum_{j=1}^L \beta^{j-1} (a_{L+1-j}^\dg - a_{L+1-j} P^\dg)$ at the right edge of the wire.  Using the fact that $P^\dg \ket{N} = \ket{N+2}$, in the limit of large $L$ and finite $p$, we find $\beta = (1-2p),$ the same decay constant as in Eq. (\ref{eq:Mj}). The minus sign in front of the annihilation operator in $\Gamma_R$ is evident for $ p = 0.5$ since the parity of permutations needed to apply $a_L^\dg$ to $\ket{N}$ is opposite from that needed to apply $a_L$ to $\ket{N+2}$.  In both expressions, we ignore the overlap of $\Gamma_{L}$ with $\Gamma_{R}$, taking the $L \to \infty$ and $\beta^L \to 0$ limits.  Including the normalization, we obtain

\beq
\Gamma^{\dg}_{L} = 4p(1-p)\sum_{j = 1}^{L } (1-2p)^{j-1} (a^{\dg}_{j} +  a_{j}P^{\dg}).\label{eq:GL}
\eeq
\beq
\Gamma^{\dg}_{R} = i\times 4p(1-p) \sum_{j = 1}^{L} (1-2p)^{j-1} (a^{\dg}_{L+1-j} -  a_{L+1-j}P^{\dg}).\label{eq:GR}
\eeq

Note that in the limit of large $N$ and $L$, $\left( \Gamma^{\dg}_{L}\right)^2 = \left(\Gamma^{\dg}_{R}\right)^2 = P^{\dg}$. Thus, while these operators do not square to unity, as the canonical Majorana operators do, they square to the operators that induce a transition between two neighboring ground states of the same parity, which is as close to the trivial operator as is possible in the number conserving case.

Given the form of the ground state wave functions, Eq. (\ref{eq:PsiN}), $\Gamma_L^\dg$ induces transitions $\ket{N}\to \ket{N+1}$, while the action of $\Gamma_R^\dg$ is more complex, $\ket{N}\to i (-1)^N\ket{N+1}$. This replicates the canonical fermionic anticommutation relations $\Gamma_L^\dg\Gamma_R^\dg = -\Gamma_R^\dg\Gamma_L^\dg$, further extending the correspondence between the $\Gamma$ operators introduced here and the Majorana operators that appear in the mean-field treatment of the Kitaev model. Note that this by itself does not imply that the operators $\Gamma$ are simple fermionic operators, only that they act as such within the ground state manifold.
Furthermore, since these operators are constrained only with respect to their action on the ground states, they may contain arbitrary terms that act in the orthogonal subspace with no visible effect for our purposes.

The $\Gamma_{L,R}$ operators given by Eqs. (\ref{eq:GL}, \ref{eq:GR}) have an explicit dependence on the number of particles $N$; however, this dependence is smooth, only via the filling $ p = N/L$. 
To make a connection with the mean-field limit, we recall that there are the number fluctuations are $O(N^{1/2})$, which translates into a variation of $p$ of order $\mathcal{O}(L^{-1/2})$ that vanishes in the large system limit. Therefore, for large systems, we can expect $\Gamma_{L,R}$ to directly correspond with the mean-field MZM operator $ \gamma$.
Indeed, $\Gamma_{L}$ and $\Gamma_{R}$ take a form closely similar to the traditional MZM operators within mean-field theory, with the annihilation operator part of the Bogoliubov quasiparticle ``decorated" by the Cooper pair operator for the wire. To convert Eqs. (\ref{eq:GL}, \ref{eq:GR}) into the mean-field expressions, it is sufficient to replace the Cooper pair creation operator with its expectation value, which is merely the superconducting order parameter. 
We finally note that the form of Majorana operators in Eqs. (\ref{eq:GL}, \ref{eq:GR}) is precisely the one conjectured by  Lin and Leggett in Ref.~\cite{lin2018towards}.

\section{Topological Entanglement Entropy}

\textcolor{black}{We can measure the topological robustness of the system whose ground state is a number-projected wave function by examining the topological entanglement entropy.} Such a measure was first proposed for two-dimensional gapped topological systems~\cite{kitaevpreskill} and is designed to isolate a constant long-range contribution to the entanglement entropy and can be related to the topological degeneracy of the ground state. Here we study a one-dimensional analog of this quantity~\cite{topoEEoned}, $\mathcal{S}_{\text{topo}},$ defined as

\begin{align}
    \mathcal{S}_{\text{topo}} &= \mathcal{S}_{AB} + \mathcal{S}_{BC} - \mathcal{S}_B - \mathcal{S}_{ABC}\label{TEE}
\end{align}
where $\mathcal{S}_{A}$ refers to, for instance, the usual von Neumann entanglement entropy of subsystem $A$. 

\begin{figure}
\includegraphics[width=3in]{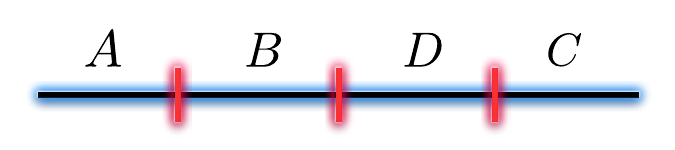}
\caption{Partitioning of the wave function into 4 regions $A, B, D, C$ to compute the topological entanglement entropy.}
\label{fig:part}
\end{figure}

For the mean-field Kitaev Hamiltonian, the behavior of $\mathcal{S}_{\text{topo}}$ has been studied in great detail, even in the presence of additional interactions, and changes abruptly from $0$ to $\log (2)$ as one enters the topological Kitaev phase from the trivial phase~\cite{topoEEoned}. For $\Delta = t$, the computation of the entanglement entropies can be carried out exactly as in~\cite{topoEEoned}; one finds $\mathcal{S}_{AB} = \mathcal{S}_B = \mathcal{S}_{ABC} = \log (2)$ while $\mathcal{S}_{BC} = 2 \log (2)$, which yields $\mathcal{S}_{\text{topo}} = \log (2)$. The $\log(2)$  implies a topological ground state degeneracy of $2$ in this case, as expected. Note that the exact size of regions $A, B, C, D$ does not affect the outcome of the calculations as one should expect for a topologically robust phase; only $\mathcal{S}_{BC}$ is different here, and this can be understood from the fact that the subsystem $\text{BC}$ is composed of disjoint parts as shown in  Fig.~\ref{fig:part}. The $\log (2)$ topological entanglement entropy can be related to the presence of two Majorana zero modes at the edge of the system. {We note crucially however that topological entanglement entropy is {\em not} in itself a consequence of the MZMs. This is most easy to see in the case $\Delta = t$. Turning on a boundary  term in the  Hamiltonian (\ref{eq:HMF}) $ \delta H_{MF } = b (a_{L}^{\dg}a_{1}  - a_{L}^{\dg}a_{1}^\dg + h.c.)$ converts from the open to the periodic boundary conditions. It is easy to check that the mean-field ground state wave function (\ref{eq:Psi})} is also an eigenstate of $\delta H_{MF}$ with the eigenvalue $(-1)^P b$; that is, the state (\ref{eq:Psi}) remains an eigenstate regardless of the boundary conditions - even though the energy of this state depends on the strength of the boundary term and the fermion parity  $P$. 
Clearly, in the case of periodic (or twisted) boundary conditions, there are no edge modes. The fact that the ground state wave function remains unchanged implies that the TEE as defined in Eq.~(\ref{TEE}) is the same for open and periodic boundary conditions. We are therefore led to conclude that this definition of TEE is sensitive to the topological properties of the state even when the system has no edges (reminiscent of the appearance of edge modes in the entanglement spectrum of topological insulators \cite{fidk2010}).

We now evaluate $\mathcal{S}_{\text{topo}}$ for the fixed-$N$ projected states and analyze its robustness compared to the mean-field limit. A straightforward calculation assuming the thermodynamic limit $N, L \rightarrow \infty$, with $p \equiv N/L, q \equiv 1-p$ (details in App.~\ref{app:TEEcalc}), yields 
\beq
 \mathcal{S}_{L_1} = \frac{1}{2} \log \left(2 \pi e p q \tilde L \right)
 \label{eq:SL}
\eeq
where $\tilde L = L_1 L_2/L$ depends on the partitioning, $L_1 + L_2 = L$, and $e$ is the Euler's constant. For entanglement entropy of disconnected segments, the expression is similar, with  $L_1$ being now the sum total of their lengths, and with an additional contribution $\log 2$.

\begin{align}
    \mathcal{S}_{AB} &= \frac{1}{2} \log \left(2 \pi e p q L_{AB} L_{CD}/L \right) \nonumber \\
    \mathcal{S}_{BC} &= \frac{1}{2} \log \left(2 \pi e p q L_{BC} L_{AD}/L \right) + \log (2) \nonumber \\
    \mathcal{S}_{B} &= \frac{1}{2} \log \left(2 \pi e p q L_{B} L_{ADC}/L \right) \nonumber \\
    \mathcal{S}_{ABC} &= \frac{1}{2} \log \left(2 \pi e p q L_{ABC} L_{D}/L \right) \nonumber \\
    \mathcal{S}_{\text{topo}} &= \frac{1}{2} \log \left( \frac{L_{AB} L_{CD} L_{BC} L_{AD}}{L_B L_{ADC} L_{ABC} L_D} \right) + \log (2)
\label{eq:topEEcalc}
\end{align}

In the limit where $L_{AB} \ll L_{CD}$, $\mathcal{S}_{AB} \approx \frac{1}{2} \log L_{AB}$; the logarithmic behavior of the entanglement entropy indicates that this wave function is the ground state of a gapless Hamiltonian \cite{hastings2007area}, which is indeed true for the Hamiltonian in Eq.~(\ref{eq:HHazzard}). This contrasts with the result for the mean-field Kitaev wave function for which the contributions to the entanglement entropy are purely area law, as the parent Hamiltonian is gapped. However, we also see that the piece $\mathcal{S}_{BC}$ has a robust geometry independent $\log (2)$ contribution which ultimately comes from the fact that $\text{BC}$ is a subsystem composed of physically disjoint parts. This $\log(2)$ agrees with the result for the mean-field Kitaev wave function and \textcolor{black}{is a signature of topological order in the projected wave function}. \textcolor{black}{We thus note that gaplessness does not preclude the presence of robust edge modes which may be algebraically or even exponentially localized~\cite{verrsengaplesstopo}. From the practical standpoint, however, it may be significantly more challenging to reach the ground state, or to perform braiding operations with such modes without exciting above the ground state}. 

\begin{figure}
\includegraphics[width=3.3in]{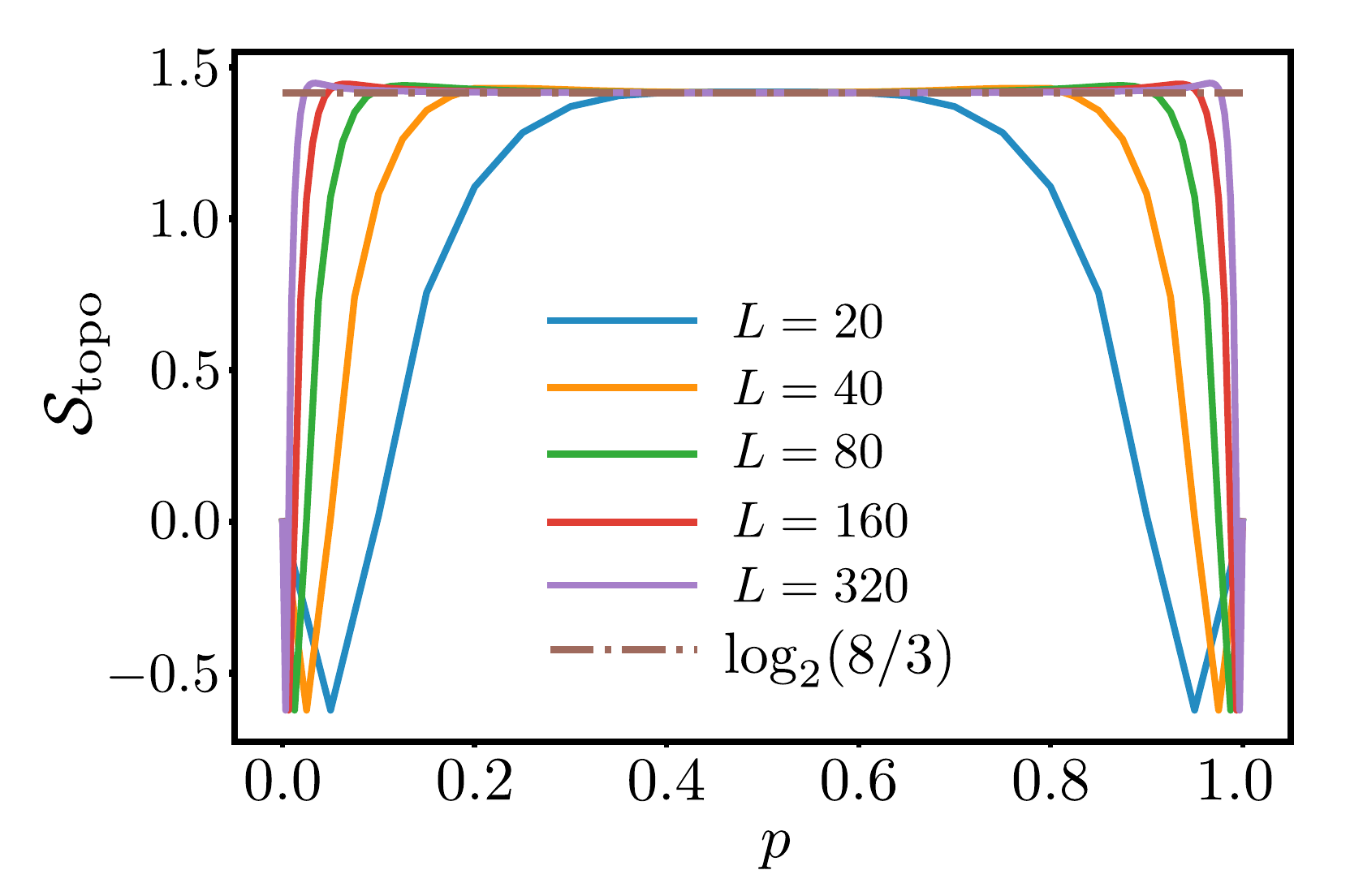}
\caption{Topological Entanglement Entropy (TEE) $S_{topo}$ plotted for various filling fractions $p.$ As we approach the infinite system limit, the TEE saturates to a value of $\log(8/3)$ for all $p \ne 0, 1.$ This is in contrast to the mean-field Kitaev limit, where the TEE is $\log(2)$ precisely in the topological phase and zero everywhere else.}
\end{figure}

To study the topological entanglement entropy more generally, we numerically compute it for arbitrary filling fraction $p = N/L$ for a particular partitioning of the system, with $L_A = L_B = L_C = L_D$, for which $\mathcal{S}_{\text{topo}} = \log (8/3)$ at arbitrary but finite $p$ [from Eq.(\ref{eq:topEEcalc})]. We find excellent numerical agreement with this result at $p = 1/2$ and find that this value is robust for a broad range of $p$ around half-filling. As the filling fraction approaches the extreme values of $p = 0, 1$, $\mathcal{S}_{\text{topo}}$ approaches $0$ as expected. However, as the system size is increased, this transition appears to become sharper. 

Finally, we note that the picture does not appear to change much when $t \neq \Delta \neq 1$ or $\mu \neq 0$, indicating the robustness of the result of Eq.~(\ref{eq:topEEcalc}). For further discussion and data to this effect, see App.~\ref{app:TEEexact}.

\section{Tunneling Conductance}
Having constructed explicit many-body operators $\Gamma_{L}^{\dg}$ and $\Gamma_{R}^{\dg}$ that induce transitions between $\ket{{N}}$ and $\ket{{N + 1}},$ we are now ready to examine whether the tunneling conductance into the edges of a wire whose ground states are given by Eq. (\ref{eq:PsiN}) differs from the well-known mean-field result of $ {2e^{2}}/{h}$ at zero bias and temperature. \cite{ngMajorana, beenakker2013search}

Without making any assumptions except that the wire has degenerate ground states $\ket{{N}}$, the coupling to a tunneling probe is described by the Hamiltonian

\bea\label{eq:tunnel}
H &=& H_{\text{lead}} + H_{\text{T}} \\&=& \sum_{k} \epsilon_{k} c^{\dg}_{k}c_{k} -  \sum_{k, N} t_{k} \ket{{N+1}}\bra{{N}}c_{k} + c_{k}^{\dg}\ket{{N}}\bra{{N+1}}\nonumber
\eea

Note that the $\sum_N'\ket{N + 1}\bra{N}$ are precisely the $\Gamma_{L}^{\dg} $ operator that we constructed earlier, assuming that we limit the range of values of $N$ to $[\bar N - M, \bar N+M]$ with $M \ll \bar N$ (signified by the prime in the summation above). \textcolor{black}{This is a standard assumption within the tunneling approximation -- that in the course of tunneling, the system's macroscopic density does not deviate significantly from its initial value $\bar N$. Here, it helps us to aggregate the tunneling terms for different $N$ into $\Gamma_L^\dagger$, which are defined for a fixed filling fraction $p$, Eq. (\ref{eq:GL}).}


To proceed, it is convenient to transform to the conjugate phase basis, $\ket{N}  = \int_0^{2\pi} \frac{d \phi}{2\pi}\, e^{iN\phi} \ket{\phi}$. In this basis, we have $\Gamma_L^\dg \propto \int_0^{2\pi}{d \phi d\phi'} e^{i\phi'} \delta_{\phi - \phi'} \ket{\phi'}\bra{\phi}$, where $\delta_x$ is the Dirac delta function with a finite width $\sim 1/M$. To capture the fermionic character of the $\Gamma$ operator, we also  introduce an auxiliary Majorana fermion mode $\gamma_L$, which plays the same role as the Klein factor in bosonization,  yielding
\bea
\Gamma_L^\dg &=& \sum_N\ket{N + 1}\bra{N} \nonumber\\
&=& \gamma_L \int_0^{2\pi}\frac{d \phi d\phi'}{4\pi^2} e^{i\phi'} \delta_{\phi - \phi'} \ket{\phi'}\bra{\phi}. \label{eq:GaL}
\eea

 This allows us to rewrite the tunneling problem Eq. (\ref{eq:tunnel}) in terms of the auxiliary Majorana operator $\gamma_L$, with the replacement of the number state representation by the phase representation \footnote{A similar expression can be derived for the other edge of the wire, 
$\Gamma_R^\dg = i\sum_N(-1)^N\ket{N + 1}\bra{N}  
= i \gamma_R \int_0^{2\pi}\frac{d \phi d\phi'}{4\pi^2} e^{i\phi'} \delta_{\phi - \phi'-\pi} \ket{\phi'}\bra{\phi}
$
},

\bea\label{eq:tunnel2}
H_T &= \int_0^{2\pi}\frac{d \phi d\phi'\, \delta_{\phi - \phi'}\ket{\phi'}\bra{\phi}}{4\pi^2}   \sum_{k} t_{k} (e^{i\phi'} c_{k} + e^{-i\phi'} c_{k}^{\dg})\gamma_L \nonumber \\ &
\eea

Note that despite the similarity with the mean-field Hamiltonian for tunneling into Majorana fermions \cite{Bolech2007}, here there is no assumption of the ordering of the superconducting phase -- the phase variable $\phi$ is a free parameter that has to be integrated over. 

To compute the tunneling conductance, we expand the operator for the tunneling current, 
\bea\label{eq:tunnelI2}
I &=& i\left[\sum_k c_{k}^{\dg} c_{k}, H_T\right] \nonumber\\ &=& i\int_0^{2\pi}\frac{d \phi d\phi'\, \delta_{\phi - \phi'}\ket{\phi'}\bra{\phi}}{4\pi^2}   \sum_{k} t_{k} (e^{i\phi'} c_{k} - e^{-i\phi'} c_{k}^{\dg})\gamma_L \nonumber. 
\eea

The expectation value of the current as a function of bias voltage defines the tunneling conductance. It can be computed using the standard methods of linear response theory \cite{mahan2000many}. 

The tunnel current, or any other observable, is computed as an expectation value over the initial state, which we can choose, for instance, to be a product state of  $\ket{N}$ in the superconductor and the filled Fermi sea in the lead, with the chemical potential different from the superconductor by the value of the applied voltage.  

All calculations of this kind are simplified by the following observation: In any order of perturbation theory, we will encounter contraction over the phase variable. Thanks to the presence of the delta function of width $1/M$, these contractions are trivial -- select a single value of phase for all terms involved, until we reach the order $\sim M$. Therefore, the result is equivalent to computing expectation values at fixed phase $\phi$, and then taking the average over it. The last step selects only the terms that are independent of $\phi$. 
Given our choice of the fixed-$N$ wave functions as the number projected version of the mean-field wave function, the tunnel current into the Majorana states is guaranteed to match the mean-field result.

We thus conclude that the tunneling conductance (or any other observable computed perturbatively over the ground state) would be the same, whether it is computed relative to the number-conserving Kitaev ground-state, or the mean-field Kitaev ground state.
In particular, we should expect the ${2e^{2}}/{h}$ zero-bias tunneling conductance result to remain unchanged. An important caveat is that if the Hamiltonian is gapless, there will be generally other contributions to conductance coming from tunneling between the lead and the other low-energy states. However, unlike the Majorana contribution, the tunneling into these states will become suppressed at weak tunneling as $t_k^2$ and thus can be filtered out.

\section{Summary and Discussion}

Motivated by the question of which properties of MZMs survive in isolated superconductors, we investigated the ground state of the Kitaev chain projected to fixed electron number sectors,  $\ket{N}$. Using the exact form of these wave functions at $\mu = 0, t = \Delta$, we were able to demonstrate the presence of zero-energy edge excitations by explicit computation of the spectral function at zero energy. The localization length of these edge modes obtained from these projected wave functions matches closely the mean-field theory of MZMs (at appropriate chemical potential) near half-filling but starts to deviate at filling fractions near the boundary of the topological and trivial phases of the mean-field Kitaev chain. Further, we constructed many-body Majorana operators in the number-conserving setting that transition between different fixed number states. These operators explicitly involve the Cooper pair operator and anticommute within the ground state manifold. Unlike the mean-field case, however, they square to the Cooper pair operator, instead of identity. We further showed that tunneling from a lead into these modified Majorana operators yields the same quantized value zero-bias tunneling conductance as in the mean-field case. 

To shed further light on the entanglement structure of the number projected wave functions $\ket{N}$, we computed their topological entanglement entropy (TEE) and found it contains a logarithmic correction to the mean-field result of $\log(2)$. This can be done analytically and exactly numerically for $t =  \Delta, \mu = 0$.  For $t \neq \Delta, \mu \neq 0$, we computed TEE by projecting the mean-field Kitaev wave functions to fixed filling fractions in a finite system; the computed TEE agrees well with that predicted from the projected mean-field $\mu = 0, t = \Delta$ wave functions, pointing at the universality of this result. 
%
However, the saturation value that we obtained in \eqref{eq:topEEcalc} contains a geometric piece that depends on the details of partitioning into subsystems $A, B, D, C$. This indicates that the Hamiltonian [such as in Eq.~(\ref{eq:HHazzard})] that realizes the wave functions $\ket{N}$ as ground states is likely to be gapless. It is yet unclear how detrimental this is to the topological robustness of the MZMs and thus requires further investigation. Whether it is possible to extract solely the partition-independent contribution to the TEE from the number-projected wave functions, e.g. by placing the system on a non-open wire geometry, is also an open question.

In this work, we focused on Hamiltonian-independent properties, which could be gleaned purely from the ground state wave function. As a result, some important questions remain outside the scope of this study. In particular, it is well established that braiding the mean-field Majorana zero-modes leads to nontrivial transformations in the degenerate ground state manifold, enabling
topological quantum computation. It remains to be seen whether a T-junction braid  \cite{alicea2011non} within a fully number-conserving regime recovers the non-Abelian statistics realized in the mean-field limit. The presence of non-topological low-energy modes may make braiding very challenging, both in theory and in practice. It is also worth considering whether a measurement-based approach to braiding could be more robust to the presence of low-energy excitations. We leave these questions to future work.

A related question is whether the procedure of computing the ground state of a mean-field Hamiltonian and projecting it to a fixed number can yield a state that is the ground state of a {\em gapped } Hamiltonian. Such a state would enable a more robust implementation of a dynamical braiding procedure and allow for a way to extract solely the universal piece of the topological entanglement entropy.

\section{Acknowledgements}
We thank Tony Leggett and Roman Lutchyn for useful discussions.  This work was funded by the Materials Sciences and Engineering Division, Basic Energy Sciences, Office of Science, US DOE.


\bibliography{refs.bib}

\begin{thebibliography}{31}%
\makeatletter
\providecommand \@ifxundefined [1]{%
 \@ifx{#1\undefined}
}%
\providecommand \@ifnum [1]{%
 \ifnum #1\expandafter \@firstoftwo
 \else \expandafter \@secondoftwo
 \fi
}%
\providecommand \@ifx [1]{%
 \ifx #1\expandafter \@firstoftwo
 \else \expandafter \@secondoftwo
 \fi
}%
\providecommand \natexlab [1]{#1}%
\providecommand \enquote  [1]{``#1''}%
\providecommand \bibnamefont  [1]{#1}%
\providecommand \bibfnamefont [1]{#1}%
\providecommand \citenamefont [1]{#1}%
\providecommand \href@noop [0]{\@secondoftwo}%
\providecommand \href [0]{\begingroup \@sanitize@url \@href}%
\providecommand \@href[1]{\@@startlink{#1}\@@href}%
\providecommand \@@href[1]{\endgroup#1\@@endlink}%
\providecommand \@sanitize@url [0]{\catcode `\\12\catcode `\$12\catcode
  `\&12\catcode `\#12\catcode `\^12\catcode `\_12\catcode `\%12\relax}%
\providecommand \@@startlink[1]{}%
\providecommand \@@endlink[0]{}%
\providecommand \url  [0]{\begingroup\@sanitize@url \@url }%
\providecommand \@url [1]{\endgroup\@href {#1}{\urlprefix }}%
\providecommand \urlprefix  [0]{URL }%
\providecommand \Eprint [0]{\href }%
\providecommand \doibase [0]{https://doi.org/}%
\providecommand \selectlanguage [0]{\@gobble}%
\providecommand \bibinfo  [0]{\@secondoftwo}%
\providecommand \bibfield  [0]{\@secondoftwo}%
\providecommand \translation [1]{[#1]}%
\providecommand \BibitemOpen [0]{}%
\providecommand \bibitemStop [0]{}%
\providecommand \bibitemNoStop [0]{.\EOS\space}%
\providecommand \EOS [0]{\spacefactor3000\relax}%
\providecommand \BibitemShut  [1]{\csname bibitem#1\endcsname}%
\let\auto@bib@innerbib\@empty
\bibitem [{\citenamefont {Emery}\ and\ \citenamefont {Kivelson}(1992)}]{2chK}%
  \BibitemOpen
  \bibfield  {author} {\bibinfo {author} {\bibfnamefont {V.~J.}\ \bibnamefont
  {Emery}}\ and\ \bibinfo {author} {\bibfnamefont {S.}~\bibnamefont
  {Kivelson}},\ }\bibfield  {title} {\bibinfo {title} {Mapping of the
  two-channel kondo problem to a resonant-level model},\ }\href
  {https://doi.org/10.1103/PhysRevB.46.10812} {\bibfield  {journal} {\bibinfo
  {journal} {Phys. Rev. B}\ }\textbf {\bibinfo {volume} {46}},\ \bibinfo
  {pages} {10812} (\bibinfo {year} {1992})}\BibitemShut {NoStop}%
\bibitem [{\citenamefont {Kitaev}(2003)}]{kitaev2003fault}%
  \BibitemOpen
  \bibfield  {author} {\bibinfo {author} {\bibfnamefont {A.~Y.}\ \bibnamefont
  {Kitaev}},\ }\bibfield  {title} {\bibinfo {title} {Fault-tolerant quantum
  computation by anyons},\ }\href@noop {} {\bibfield  {journal} {\bibinfo
  {journal} {Annals of physics}\ }\textbf {\bibinfo {volume} {303}},\ \bibinfo
  {pages} {2} (\bibinfo {year} {2003})}\BibitemShut {NoStop}%
\bibitem [{\citenamefont {Kitaev}(2006)}]{kitaev2006anyons}%
  \BibitemOpen
  \bibfield  {author} {\bibinfo {author} {\bibfnamefont {A.}~\bibnamefont
  {Kitaev}},\ }\bibfield  {title} {\bibinfo {title} {Anyons in an exactly
  solved model and beyond},\ }\href@noop {} {\bibfield  {journal} {\bibinfo
  {journal} {Annals of Physics}\ }\textbf {\bibinfo {volume} {321}},\ \bibinfo
  {pages} {2} (\bibinfo {year} {2006})}\BibitemShut {NoStop}%
\bibitem [{\citenamefont {Kitaev}(2001)}]{kitaev2001unpaired}%
  \BibitemOpen
  \bibfield  {author} {\bibinfo {author} {\bibfnamefont {A.~Y.}\ \bibnamefont
  {Kitaev}},\ }\bibfield  {title} {\bibinfo {title} {Unpaired majorana fermions
  in quantum wires},\ }\href@noop {} {\bibfield  {journal} {\bibinfo  {journal}
  {Physics-uspekhi}\ }\textbf {\bibinfo {volume} {44}},\ \bibinfo {pages} {131}
  (\bibinfo {year} {2001})}\BibitemShut {NoStop}%
\bibitem [{\citenamefont {Read}\ and\ \citenamefont
  {Green}(2000)}]{read2000paired}%
  \BibitemOpen
  \bibfield  {author} {\bibinfo {author} {\bibfnamefont {N.}~\bibnamefont
  {Read}}\ and\ \bibinfo {author} {\bibfnamefont {D.}~\bibnamefont {Green}},\
  }\bibfield  {title} {\bibinfo {title} {Paired states of fermions in two
  dimensions with breaking of parity and time-reversal symmetries and the
  fractional quantum hall effect},\ }\href@noop {} {\bibfield  {journal}
  {\bibinfo  {journal} {Physical Review B}\ }\textbf {\bibinfo {volume} {61}},\
  \bibinfo {pages} {10267} (\bibinfo {year} {2000})}\BibitemShut {NoStop}%
\bibitem [{\citenamefont {Ivanov}(2001)}]{ivanov2001non}%
  \BibitemOpen
  \bibfield  {author} {\bibinfo {author} {\bibfnamefont {D.~A.}\ \bibnamefont
  {Ivanov}},\ }\bibfield  {title} {\bibinfo {title} {Non-abelian statistics of
  half-quantum vortices in p-wave superconductors},\ }\href@noop {} {\bibfield
  {journal} {\bibinfo  {journal} {Physical review letters}\ }\textbf {\bibinfo
  {volume} {86}},\ \bibinfo {pages} {268} (\bibinfo {year} {2001})}\BibitemShut
  {NoStop}%
\bibitem [{\citenamefont {Nayak}\ \emph {et~al.}(2008)\citenamefont {Nayak},
  \citenamefont {Simon}, \citenamefont {Stern}, \citenamefont {Freedman},\ and\
  \citenamefont {Sarma}}]{nayak2008non}%
  \BibitemOpen
  \bibfield  {author} {\bibinfo {author} {\bibfnamefont {C.}~\bibnamefont
  {Nayak}}, \bibinfo {author} {\bibfnamefont {S.~H.}\ \bibnamefont {Simon}},
  \bibinfo {author} {\bibfnamefont {A.}~\bibnamefont {Stern}}, \bibinfo
  {author} {\bibfnamefont {M.}~\bibnamefont {Freedman}},\ and\ \bibinfo
  {author} {\bibfnamefont {S.~D.}\ \bibnamefont {Sarma}},\ }\bibfield  {title}
  {\bibinfo {title} {Non-abelian anyons and topological quantum computation},\
  }\href@noop {} {\bibfield  {journal} {\bibinfo  {journal} {Reviews of Modern
  Physics}\ }\textbf {\bibinfo {volume} {80}},\ \bibinfo {pages} {1083}
  (\bibinfo {year} {2008})}\BibitemShut {NoStop}%
\bibitem [{\citenamefont {Lin}\ and\ \citenamefont
  {Leggett}(2018)}]{lin2018towards}%
  \BibitemOpen
  \bibfield  {author} {\bibinfo {author} {\bibfnamefont {Y.}~\bibnamefont
  {Lin}}\ and\ \bibinfo {author} {\bibfnamefont {A.~J.}\ \bibnamefont
  {Leggett}},\ }\bibfield  {title} {\bibinfo {title} {Towards a particle-number
  conserving theory of majorana zero modes in p+ ip superfluids},\ }\href@noop
  {} {\bibfield  {journal} {\bibinfo  {journal} {arXiv preprint
  arXiv:1803.08003}\ } (\bibinfo {year} {2018})}\BibitemShut {NoStop}%
\bibitem [{\citenamefont {Lutchyn}\ \emph {et~al.}(2018)\citenamefont
  {Lutchyn}, \citenamefont {Bakkers}, \citenamefont {Kouwenhoven},
  \citenamefont {Krogstrup}, \citenamefont {Marcus},\ and\ \citenamefont
  {Oreg}}]{lutchyn2018majorana}%
  \BibitemOpen
  \bibfield  {author} {\bibinfo {author} {\bibfnamefont {R.~M.}\ \bibnamefont
  {Lutchyn}}, \bibinfo {author} {\bibfnamefont {E.~P.}\ \bibnamefont
  {Bakkers}}, \bibinfo {author} {\bibfnamefont {L.~P.}\ \bibnamefont
  {Kouwenhoven}}, \bibinfo {author} {\bibfnamefont {P.}~\bibnamefont
  {Krogstrup}}, \bibinfo {author} {\bibfnamefont {C.~M.}\ \bibnamefont
  {Marcus}},\ and\ \bibinfo {author} {\bibfnamefont {Y.}~\bibnamefont {Oreg}},\
  }\bibfield  {title} {\bibinfo {title} {Majorana zero modes in
  superconductor--semiconductor heterostructures},\ }\href@noop {} {\bibfield
  {journal} {\bibinfo  {journal} {Nature Reviews Materials}\ }\textbf {\bibinfo
  {volume} {3}},\ \bibinfo {pages} {52} (\bibinfo {year} {2018})}\BibitemShut
  {NoStop}%
\bibitem [{\citenamefont {Leggett}(2006)}]{leggett2006quantum}%
  \BibitemOpen
  \bibfield  {author} {\bibinfo {author} {\bibfnamefont {A.~J.}\ \bibnamefont
  {Leggett}},\ }\href@noop {} {\emph {\bibinfo {title} {Quantum liquids: Bose
  condensation and Cooper pairing in condensed-matter systems}}}\ (\bibinfo
  {publisher} {Oxford university press},\ \bibinfo {year} {2006})\BibitemShut
  {NoStop}%
\bibitem [{\citenamefont {Iemini}\ \emph {et~al.}(2015)\citenamefont {Iemini},
  \citenamefont {Mazza}, \citenamefont {Rossini}, \citenamefont {Fazio},\ and\
  \citenamefont {Diehl}}]{Iemini15}%
  \BibitemOpen
  \bibfield  {author} {\bibinfo {author} {\bibfnamefont {F.}~\bibnamefont
  {Iemini}}, \bibinfo {author} {\bibfnamefont {L.}~\bibnamefont {Mazza}},
  \bibinfo {author} {\bibfnamefont {D.}~\bibnamefont {Rossini}}, \bibinfo
  {author} {\bibfnamefont {R.}~\bibnamefont {Fazio}},\ and\ \bibinfo {author}
  {\bibfnamefont {S.}~\bibnamefont {Diehl}},\ }\bibfield  {title} {\bibinfo
  {title} {Localized majorana-like modes in a number-conserving setting: An
  exactly solvable model},\ }\href
  {https://doi.org/10.1103/PhysRevLett.115.156402} {\bibfield  {journal}
  {\bibinfo  {journal} {Phys. Rev. Lett.}\ }\textbf {\bibinfo {volume} {115}},\
  \bibinfo {pages} {156402} (\bibinfo {year} {2015})}\BibitemShut {NoStop}%
\bibitem [{\citenamefont {Wang}\ \emph {et~al.}(2017)\citenamefont {Wang},
  \citenamefont {Xu}, \citenamefont {Pu},\ and\ \citenamefont
  {Hazzard}}]{WangNumberConserving}%
  \BibitemOpen
  \bibfield  {author} {\bibinfo {author} {\bibfnamefont {Z.}~\bibnamefont
  {Wang}}, \bibinfo {author} {\bibfnamefont {Y.}~\bibnamefont {Xu}}, \bibinfo
  {author} {\bibfnamefont {H.}~\bibnamefont {Pu}},\ and\ \bibinfo {author}
  {\bibfnamefont {K.~R.~A.}\ \bibnamefont {Hazzard}},\ }\bibfield  {title}
  {\bibinfo {title} {Number-conserving interacting fermion models with exact
  topological superconducting ground states},\ }\href
  {https://doi.org/10.1103/PhysRevB.96.115110} {\bibfield  {journal} {\bibinfo
  {journal} {Phys. Rev. B}\ }\textbf {\bibinfo {volume} {96}},\ \bibinfo
  {pages} {115110} (\bibinfo {year} {2017})}\BibitemShut {NoStop}%
\bibitem [{\citenamefont {Ortiz}\ \emph {et~al.}(2014)\citenamefont {Ortiz},
  \citenamefont {Dukelsky}, \citenamefont {Cobanera}, \citenamefont {Esebbag},\
  and\ \citenamefont {Beenakker}}]{Ortiz14}%
  \BibitemOpen
  \bibfield  {author} {\bibinfo {author} {\bibfnamefont {G.}~\bibnamefont
  {Ortiz}}, \bibinfo {author} {\bibfnamefont {J.}~\bibnamefont {Dukelsky}},
  \bibinfo {author} {\bibfnamefont {E.}~\bibnamefont {Cobanera}}, \bibinfo
  {author} {\bibfnamefont {C.}~\bibnamefont {Esebbag}},\ and\ \bibinfo {author}
  {\bibfnamefont {C.}~\bibnamefont {Beenakker}},\ }\bibfield  {title} {\bibinfo
  {title} {Many-body characterization of particle-conserving topological
  superfluids},\ }\href {https://doi.org/10.1103/PhysRevLett.113.267002}
  {\bibfield  {journal} {\bibinfo  {journal} {Phys. Rev. Lett.}\ }\textbf
  {\bibinfo {volume} {113}},\ \bibinfo {pages} {267002} (\bibinfo {year}
  {2014})}\BibitemShut {NoStop}%
\bibitem [{\citenamefont {Lang}\ and\ \citenamefont
  {B\"uchler}(2015)}]{Lang2015}%
  \BibitemOpen
  \bibfield  {author} {\bibinfo {author} {\bibfnamefont {N.}~\bibnamefont
  {Lang}}\ and\ \bibinfo {author} {\bibfnamefont {H.~P.}\ \bibnamefont
  {B\"uchler}},\ }\bibfield  {title} {\bibinfo {title} {Topological states in a
  microscopic model of interacting fermions},\ }\href
  {https://doi.org/10.1103/PhysRevB.92.041118} {\bibfield  {journal} {\bibinfo
  {journal} {Phys. Rev. B}\ }\textbf {\bibinfo {volume} {92}},\ \bibinfo
  {pages} {041118} (\bibinfo {year} {2015})}\BibitemShut {NoStop}%
\bibitem [{\citenamefont {Sau}\ \emph {et~al.}(2011)\citenamefont {Sau},
  \citenamefont {Halperin}, \citenamefont {Flensberg},\ and\ \citenamefont
  {Das~Sarma}}]{Sau2011}%
  \BibitemOpen
  \bibfield  {author} {\bibinfo {author} {\bibfnamefont {J.~D.}\ \bibnamefont
  {Sau}}, \bibinfo {author} {\bibfnamefont {B.~I.}\ \bibnamefont {Halperin}},
  \bibinfo {author} {\bibfnamefont {K.}~\bibnamefont {Flensberg}},\ and\
  \bibinfo {author} {\bibfnamefont {S.}~\bibnamefont {Das~Sarma}},\ }\bibfield
  {title} {\bibinfo {title} {Number conserving theory for topologically
  protected degeneracy in one-dimensional fermions},\ }\href
  {https://doi.org/10.1103/PhysRevB.84.144509} {\bibfield  {journal} {\bibinfo
  {journal} {Phys. Rev. B}\ }\textbf {\bibinfo {volume} {84}},\ \bibinfo
  {pages} {144509} (\bibinfo {year} {2011})}\BibitemShut {NoStop}%
\bibitem [{\citenamefont {Cheng}\ and\ \citenamefont {Tu}(2011)}]{Cheng11}%
  \BibitemOpen
  \bibfield  {author} {\bibinfo {author} {\bibfnamefont {M.}~\bibnamefont
  {Cheng}}\ and\ \bibinfo {author} {\bibfnamefont {H.-H.}\ \bibnamefont {Tu}},\
  }\bibfield  {title} {\bibinfo {title} {Majorana edge states in interacting
  two-chain ladders of fermions},\ }\href
  {https://doi.org/10.1103/PhysRevB.84.094503} {\bibfield  {journal} {\bibinfo
  {journal} {Phys. Rev. B}\ }\textbf {\bibinfo {volume} {84}},\ \bibinfo
  {pages} {094503} (\bibinfo {year} {2011})}\BibitemShut {NoStop}%
\bibitem [{\citenamefont {Fidkowski}\ \emph {et~al.}(2011)\citenamefont
  {Fidkowski}, \citenamefont {Lutchyn}, \citenamefont {Nayak},\ and\
  \citenamefont {Fisher}}]{Lukasz2011}%
  \BibitemOpen
  \bibfield  {author} {\bibinfo {author} {\bibfnamefont {L.}~\bibnamefont
  {Fidkowski}}, \bibinfo {author} {\bibfnamefont {R.~M.}\ \bibnamefont
  {Lutchyn}}, \bibinfo {author} {\bibfnamefont {C.}~\bibnamefont {Nayak}},\
  and\ \bibinfo {author} {\bibfnamefont {M.~P.~A.}\ \bibnamefont {Fisher}},\
  }\bibfield  {title} {\bibinfo {title} {Majorana zero modes in one-dimensional
  quantum wires without long-ranged superconducting order},\ }\href
  {https://doi.org/10.1103/PhysRevB.84.195436} {\bibfield  {journal} {\bibinfo
  {journal} {Phys. Rev. B}\ }\textbf {\bibinfo {volume} {84}},\ \bibinfo
  {pages} {195436} (\bibinfo {year} {2011})}\BibitemShut {NoStop}%
\bibitem [{\citenamefont {Ruhman}\ \emph {et~al.}(2015)\citenamefont {Ruhman},
  \citenamefont {Berg},\ and\ \citenamefont {Altman}}]{Ruhman2015}%
  \BibitemOpen
  \bibfield  {author} {\bibinfo {author} {\bibfnamefont {J.}~\bibnamefont
  {Ruhman}}, \bibinfo {author} {\bibfnamefont {E.}~\bibnamefont {Berg}},\ and\
  \bibinfo {author} {\bibfnamefont {E.}~\bibnamefont {Altman}},\ }\bibfield
  {title} {\bibinfo {title} {Topological states in a one-dimensional fermi gas
  with attractive interaction},\ }\href
  {https://doi.org/10.1103/PhysRevLett.114.100401} {\bibfield  {journal}
  {\bibinfo  {journal} {Phys. Rev. Lett.}\ }\textbf {\bibinfo {volume} {114}},\
  \bibinfo {pages} {100401} (\bibinfo {year} {2015})}\BibitemShut {NoStop}%
\bibitem [{\citenamefont {Kraus}\ \emph {et~al.}(2013)\citenamefont {Kraus},
  \citenamefont {Dalmonte}, \citenamefont {Baranov}, \citenamefont
  {L\"auchli},\ and\ \citenamefont {Zoller}}]{Kraus2013}%
  \BibitemOpen
  \bibfield  {author} {\bibinfo {author} {\bibfnamefont {C.~V.}\ \bibnamefont
  {Kraus}}, \bibinfo {author} {\bibfnamefont {M.}~\bibnamefont {Dalmonte}},
  \bibinfo {author} {\bibfnamefont {M.~A.}\ \bibnamefont {Baranov}}, \bibinfo
  {author} {\bibfnamefont {A.~M.}\ \bibnamefont {L\"auchli}},\ and\ \bibinfo
  {author} {\bibfnamefont {P.}~\bibnamefont {Zoller}},\ }\bibfield  {title}
  {\bibinfo {title} {Majorana edge states in atomic wires coupled by pair
  hopping},\ }\href {https://doi.org/10.1103/PhysRevLett.111.173004} {\bibfield
   {journal} {\bibinfo  {journal} {Phys. Rev. Lett.}\ }\textbf {\bibinfo
  {volume} {111}},\ \bibinfo {pages} {173004} (\bibinfo {year}
  {2013})}\BibitemShut {NoStop}%
\bibitem [{\citenamefont {Keselman}\ and\ \citenamefont
  {Berg}(2015)}]{Keselman2015}%
  \BibitemOpen
  \bibfield  {author} {\bibinfo {author} {\bibfnamefont {A.}~\bibnamefont
  {Keselman}}\ and\ \bibinfo {author} {\bibfnamefont {E.}~\bibnamefont
  {Berg}},\ }\bibfield  {title} {\bibinfo {title} {Gapless symmetry-protected
  topological phase of fermions in one dimension},\ }\href
  {https://doi.org/10.1103/PhysRevB.91.235309} {\bibfield  {journal} {\bibinfo
  {journal} {Phys. Rev. B}\ }\textbf {\bibinfo {volume} {91}},\ \bibinfo
  {pages} {235309} (\bibinfo {year} {2015})}\BibitemShut {NoStop}%
\bibitem [{\citenamefont {Hastings}(2007)}]{hastings2007area}%
  \BibitemOpen
  \bibfield  {author} {\bibinfo {author} {\bibfnamefont {M.~B.}\ \bibnamefont
  {Hastings}},\ }\bibfield  {title} {\bibinfo {title} {An area law for
  one-dimensional quantum systems},\ }\href@noop {} {\bibfield  {journal}
  {\bibinfo  {journal} {Journal of statistical mechanics: theory and
  experiment}\ }\textbf {\bibinfo {volume} {2007}},\ \bibinfo {pages} {P08024}
  (\bibinfo {year} {2007})}\BibitemShut {NoStop}%
\bibitem [{\citenamefont {Verresen}\ \emph {et~al.}(2021)\citenamefont
  {Verresen}, \citenamefont {Thorngren}, \citenamefont {Jones},\ and\
  \citenamefont {Pollmann}}]{verrsengaplesstopo}%
  \BibitemOpen
  \bibfield  {author} {\bibinfo {author} {\bibfnamefont {R.}~\bibnamefont
  {Verresen}}, \bibinfo {author} {\bibfnamefont {R.}~\bibnamefont {Thorngren}},
  \bibinfo {author} {\bibfnamefont {N.~G.}\ \bibnamefont {Jones}},\ and\
  \bibinfo {author} {\bibfnamefont {F.}~\bibnamefont {Pollmann}},\ }\bibfield
  {title} {\bibinfo {title} {Gapless topological phases and symmetry-enriched
  quantum criticality},\ }\href {https://doi.org/10.1103/PhysRevX.11.041059}
  {\bibfield  {journal} {\bibinfo  {journal} {Phys. Rev. X}\ }\textbf {\bibinfo
  {volume} {11}},\ \bibinfo {pages} {041059} (\bibinfo {year}
  {2021})}\BibitemShut {NoStop}%
\bibitem [{\citenamefont {Alicea}\ \emph {et~al.}(2011)\citenamefont {Alicea},
  \citenamefont {Oreg}, \citenamefont {Refael}, \citenamefont {Von~Oppen},\
  and\ \citenamefont {Fisher}}]{alicea2011non}%
  \BibitemOpen
  \bibfield  {author} {\bibinfo {author} {\bibfnamefont {J.}~\bibnamefont
  {Alicea}}, \bibinfo {author} {\bibfnamefont {Y.}~\bibnamefont {Oreg}},
  \bibinfo {author} {\bibfnamefont {G.}~\bibnamefont {Refael}}, \bibinfo
  {author} {\bibfnamefont {F.}~\bibnamefont {Von~Oppen}},\ and\ \bibinfo
  {author} {\bibfnamefont {M.~P.}\ \bibnamefont {Fisher}},\ }\bibfield  {title}
  {\bibinfo {title} {Non-abelian statistics and topological quantum information
  processing in 1d wire networks},\ }\href@noop {} {\bibfield  {journal}
  {\bibinfo  {journal} {Nature Physics}\ }\textbf {\bibinfo {volume} {7}},\
  \bibinfo {pages} {412} (\bibinfo {year} {2011})}\BibitemShut {NoStop}%
\bibitem [{\citenamefont {Kitaev}\ and\ \citenamefont
  {Preskill}(2006)}]{kitaevpreskill}%
  \BibitemOpen
  \bibfield  {author} {\bibinfo {author} {\bibfnamefont {A.}~\bibnamefont
  {Kitaev}}\ and\ \bibinfo {author} {\bibfnamefont {J.}~\bibnamefont
  {Preskill}},\ }\bibfield  {title} {\bibinfo {title} {Topological entanglement
  entropy},\ }\href {https://doi.org/10.1103/PhysRevLett.96.110404} {\bibfield
  {journal} {\bibinfo  {journal} {Phys. Rev. Lett.}\ }\textbf {\bibinfo
  {volume} {96}},\ \bibinfo {pages} {110404} (\bibinfo {year}
  {2006})}\BibitemShut {NoStop}%
\bibitem [{\citenamefont {Fromholz}\ \emph {et~al.}(2020)\citenamefont
  {Fromholz}, \citenamefont {Magnifico}, \citenamefont {Vitale}, \citenamefont
  {Mendes-Santos},\ and\ \citenamefont {Dalmonte}}]{topoEEoned}%
  \BibitemOpen
  \bibfield  {author} {\bibinfo {author} {\bibfnamefont {P.}~\bibnamefont
  {Fromholz}}, \bibinfo {author} {\bibfnamefont {G.}~\bibnamefont {Magnifico}},
  \bibinfo {author} {\bibfnamefont {V.}~\bibnamefont {Vitale}}, \bibinfo
  {author} {\bibfnamefont {T.}~\bibnamefont {Mendes-Santos}},\ and\ \bibinfo
  {author} {\bibfnamefont {M.}~\bibnamefont {Dalmonte}},\ }\bibfield  {title}
  {\bibinfo {title} {Entanglement topological invariants for one-dimensional
  topological superconductors},\ }\href
  {https://doi.org/10.1103/PhysRevB.101.085136} {\bibfield  {journal} {\bibinfo
   {journal} {Phys. Rev. B}\ }\textbf {\bibinfo {volume} {101}},\ \bibinfo
  {pages} {085136} (\bibinfo {year} {2020})}\BibitemShut {NoStop}%
\bibitem [{\citenamefont {Fidkowski}(2010)}]{fidk2010}%
  \BibitemOpen
  \bibfield  {author} {\bibinfo {author} {\bibfnamefont {L.}~\bibnamefont
  {Fidkowski}},\ }\bibfield  {title} {\bibinfo {title} {Entanglement spectrum
  of topological insulators and superconductors},\ }\href
  {https://doi.org/10.1103/PhysRevLett.104.130502} {\bibfield  {journal}
  {\bibinfo  {journal} {Phys. Rev. Lett.}\ }\textbf {\bibinfo {volume} {104}},\
  \bibinfo {pages} {130502} (\bibinfo {year} {2010})}\BibitemShut {NoStop}%
\bibitem [{\citenamefont {Law}\ \emph {et~al.}(2009)\citenamefont {Law},
  \citenamefont {Lee},\ and\ \citenamefont {Ng}}]{ngMajorana}%
  \BibitemOpen
  \bibfield  {author} {\bibinfo {author} {\bibfnamefont {K.~T.}\ \bibnamefont
  {Law}}, \bibinfo {author} {\bibfnamefont {P.~A.}\ \bibnamefont {Lee}},\ and\
  \bibinfo {author} {\bibfnamefont {T.~K.}\ \bibnamefont {Ng}},\ }\bibfield
  {title} {\bibinfo {title} {Majorana fermion induced resonant andreev
  reflection},\ }\href {https://doi.org/10.1103/PhysRevLett.103.237001}
  {\bibfield  {journal} {\bibinfo  {journal} {Phys. Rev. Lett.}\ }\textbf
  {\bibinfo {volume} {103}},\ \bibinfo {pages} {237001} (\bibinfo {year}
  {2009})}\BibitemShut {NoStop}%
\bibitem [{\citenamefont {Beenakker}(2013)}]{beenakker2013search}%
  \BibitemOpen
  \bibfield  {author} {\bibinfo {author} {\bibfnamefont {C.}~\bibnamefont
  {Beenakker}},\ }\bibfield  {title} {\bibinfo {title} {Search for majorana
  fermions in superconductors},\ }\href@noop {} {\bibfield  {journal} {\bibinfo
   {journal} {Annu. Rev. Condens. Matter Phys.}\ }\textbf {\bibinfo {volume}
  {4}},\ \bibinfo {pages} {113} (\bibinfo {year} {2013})}\BibitemShut {NoStop}%
\bibitem [{Note1()}]{Note1}%
  \BibitemOpen
  \bibinfo {note} {A similar expression can be derived for the other edge of
  the wire, $\Gamma _R^\protect \ensuremath {\dagger }= i\DOTSB \sum@ \slimits@
  _N(-1)^N\left | N + 1 \right >\left < N \right | = i \gamma _R \DOTSI \intop
  \ilimits@ _0^{2\pi }\protect \frac {d \phi d\phi '}{4\pi ^2} e^{i\phi '}
  \delta _{\phi - \phi '-\pi } \left | \phi ' \right >\left < \phi \right |
  $}\BibitemShut {NoStop}%
\bibitem [{\citenamefont {Bolech}\ and\ \citenamefont
  {Demler}(2007)}]{Bolech2007}%
  \BibitemOpen
  \bibfield  {author} {\bibinfo {author} {\bibfnamefont {C.~J.}\ \bibnamefont
  {Bolech}}\ and\ \bibinfo {author} {\bibfnamefont {E.}~\bibnamefont
  {Demler}},\ }\bibfield  {title} {\bibinfo {title} {Observing majorana bound
  states in $p$-wave superconductors using noise measurements in tunneling
  experiments},\ }\href {https://doi.org/10.1103/PhysRevLett.98.237002}
  {\bibfield  {journal} {\bibinfo  {journal} {Phys. Rev. Lett.}\ }\textbf
  {\bibinfo {volume} {98}},\ \bibinfo {pages} {237002} (\bibinfo {year}
  {2007})}\BibitemShut {NoStop}%
\bibitem [{\citenamefont {Mahan}(2000)}]{mahan2000many}%
  \BibitemOpen
  \bibfield  {author} {\bibinfo {author} {\bibfnamefont {G.~D.}\ \bibnamefont
  {Mahan}},\ }\href@noop {} {\emph {\bibinfo {title} {Many-particle physics}}}\
  (\bibinfo  {publisher} {Springer Science \& Business Media},\ \bibinfo {year}
  {2000})\BibitemShut {NoStop}%
\end{thebibliography}%

\appendix

\section{Calculation of topological entanglement entropy}
\label{app:TEEcalc}

\subsection{Numerical computation of the entanglement entropy at arbitrary filling}

The wave function of interest is a linear superposition of all possible bitstrings that satisfy the constraint on the total particle number, Eq. (\ref{eq:PsiN}). Thus, 

\begin{align}
    \ket{\psi_N} = \frac{1}{\sqrt{L\choose N}} \sum_{\{ABDC\}} \ket{ABDC}
\end{align}
where $\{ABDC\}$ represent all possible configurations of particles in the entire system satisfying the total particle number constraint. To evaluate traces over subsystems, it is useful to divide the Hilbert space into tensor product of the subspace being traced out and its complement. To do so, it is necessary to take care of the fermionic sign factors. For instance, 

\begin{align}
    \ket{ABDC} &= \ket{AB} \otimes \ket{DC}, \nonumber \\
    \ket{ABDC} &= \left(-1 \right)^{N_A N_B} \ket{B} \otimes \ket{ADC}, \nonumber \\
    \ket{ABDC} &= \left(-1 \right)^{N_D N_B} \ket{AD} \otimes \ket{BC}. 
\end{align}

Then, we can evaluate $\mathcal{S}_{AB}$ as follows. 

\begin{align}
    \rho_{AB} &= \sum_{\{DC\}} \braket{DC}{\psi_N} \braket{\psi_N}{DC} \nonumber \\
              &= \frac{1}{ {L \choose N}} \sum_{\substack{\{DC\} \\ \{AB\},\{AB'\} \\ N_{AB} = N_{AB'} = N' \\ N_{DC} = N-N'}} \ket{AB} \bra{AB'} \nonumber \\
              &= \frac{1}{ {L \choose N}} \sum_{\substack{\{DC\} \\ N_{DC} = N-N'}} \left( \sum_{\substack{\{AB\} \\ N_{AB} = N'}} \ket{AB} \right) \left( \sum_{\substack{\{AB'\} \\ N_{AB'} = N'}} \bra{AB'} \right) \nonumber \\
              &= \sum_{N'} \frac{{L_{DC} \choose {N-N'}} {L_{AB} \choose {N'}} }{{L \choose N} } \ket{\psi^{AB}_{N'}} \bra{\psi^{AB}_{N'}} \nonumber \\
              &= \sum_{N'} p_{N'} \ket{\psi^{AB}_{N'}} \bra{\psi^{AB}_{N'}} 
\end{align}
Here, $\ket{\psi^{AB}_{N'}}$ are orthonormal  wave functions (for different $N'$) that are defined on the $AB$ subsystem. The density matrix is thus already diagonalized and the entanglement entropy can be computed directly as $\mathcal{S}_{AB} = - \sum_{N'} p_{N'} \log (p_{N'})$. 

Similar considerations apply for computing the entanglement entropies $\mathcal{S}_{B}, \mathcal{S}_{ABC} = \mathcal{S}_D$ which are effectively composed of just a single contiguous subsystem. 

The computation of $\rho_{BC}$ is a bit more involved. We find

\begin{align}
    \rho_{BC} &= \frac{1}{ {L \choose N}} \sum_{\substack{\{AD\} \\ \{BC\},\{BC'\} \\ N_{BC} = N_{BC'} = N' \\ N_{AD} = N-N'}} \left( -1 \right)^{N_D (N_B + N_{B'})} \ket{BC} \bra{BC'} \nonumber \\
              &= \frac{1}{ {L \choose N}} \sum_{\substack{\{AD\} \\ N_{AD} = N-N'}} \left( \sum_{\substack{\{BC\} \\ N_{BC} = N'}} \left( -1 \right)^{N_D N_B}\ket{BC} \right) \times \nonumber \\
              & \hspace{3.5cm} \left( \sum_{\substack{\{BC'\} \\ N_{BC'} = N'}} \left( -1 \right)^{N_D N_{B'}} \bra{BC'} \right) \nonumber \\
              &= \frac{1}{ {L \choose N}} \sum_{N'} \Bigg( \sum_{\substack{\{AD\} \\ N_D = \text{even}}} \ket{\psi^{BC}_{N',+}} \bra{\psi^{BC}_{N',+}} \nonumber \\
              & \hspace{3.5cm} + \sum_{\substack{\{AD\} \\ N_D = \text{odd}}} \ket{\psi^{BC}_{N',-}} \bra{\psi^{BC}_{N',-}} \Bigg)
\end{align}

Here, 
\begin{align}
    \ket{\psi^{BC}_{N',+}}&= \sum_{\substack{ \{BC\} \\ N_{BC} = N'}} \ket{BC} \nonumber \\
    \ket{\psi^{BC}_{N',-}} &= \sum_{\substack{ \{BC\} \\ N_{BC} = N'}} (-1)^{N_B} \ket{BC} \nonumber \\
\end{align}
are unnormalized wave functions that are orthogonal for different $N'$ but not orthogonal to one another at the same $N'$. This is crucial in obtaining the extra $\log(2)$ entropy in $\mathcal{S}_{BC}$. These wave functions are easy to normalize by themselves  by simply counting the number of distinct configurations $\{BC\}$ with the number of particles $N'$; the sums $\sum_{\substack{\{AD\} \\ N_D = \text{odd/even}}}$ can then be replaced by appropriate sums over binomial coefficients. If one notes that the two wave functions $\ket{\psi^{BC}_{N',+}}, \ket{\psi^{BC}_{N',-}}$ are approximately orthogonal in the thermodynamic limit, then the presence of the extra $\log(2)$ factor becomes evident -- the probability of a state with $N'$ particles, $p_{N'}$ as evaluated in $\rho_{AB}$ is effectively replaced by $p_{N'}/2$ for two orthogonal states with the same total number of particles. Thus, the entanglement entropy changes from $-p_{N'} \log (p_{N'})$ to $-2 \times p_{N'}/2 \log (p_{N'}/2) = - p_{N'} \log (p_{N'}) - p_{N'} \log (2)$. Summing over all $N'$, the second term yields the extra $\log(2)$ entropy. 

More accurately, since $\ket{\psi^{BC}_{N',+}}, \ket{\psi^{BC}_{N',-}}$ are not precisely orthogonal in a finite system, we evaluate $\mathcal{S}_{BC}$ numerically by diagonalizing two-by-two sectors of the density matrix and computing the von-Neumann entropy with the eigenvalues of this diagonalized density matrix in the usual way. 

\subsection{Analytical calculation of the (topological) entanglement entropy in the thermodynamic limit}

The entanglement entropies noted above can be analytically computed in the thermodynamic limit using the following argument. In general, we need to compute the entanglement entropy for a probability distribution with weights

\begin{align}
    p (n_1, n_2) &= \frac{{l_1 \choose n_1} {l_2 \choose n_2}}{{l \choose n}}
\end{align}

and for which $n_1 + n_2 = n, l_1 + l_2 = l$. Assuming we are at a filling fraction $p$ (and $q = 1-p$), we can interpret the combinatorial factors as arising from a Binomial distribution and relate the result, in the thermodynamic limit, to an appropriate Gaussian distribution, using the Central Limit Theorem. In particular, using 

\begin{align}
    {l \choose n} p^n q^{l-n} \approx \frac{1}{\sqrt{2 \pi p q l}} e^{-\frac{(n - p l)^2}{2 p q l}}, 
\end{align}

and a bit of massaging, we obtain the result

\begin{align}
    p (n_1, n_2) &\approx \frac{1}{\sqrt{2 \pi p q l_1 l_2 / l}} e^{-\frac{(n_1 - p l_1)^2}{2 p q l_1 l_2 / l}}. 
\end{align}

Thus, the weights of the probability distribution of interest can be interpreted as arising from a Gaussian of width $\sigma^2 = p q l_1 l_2 / l$, and mean $\mu = p l_1$. 

It is simple to show that the von Neumann entropy associated with a Gaussian random variable is given by $\mathcal{S}_{\text{Gaussian}} = \frac{1}{2} \text{log} \left( 2 \pi e \sigma^2 \right)$. Using the value $\sigma^2 = p q l_1 l_2 / l$ for the distribution of interest, we obtain the result quoted in the main text in Eq.~(\ref{eq:SL}). The entropies of contiguous blocks can be obtained as a direct application of this result, and the entanglement entropy of the discontiguous block, $\mathcal{S}_{BC}$, can be obtained by noting the extra factor of $2$ degeneracy in the entanglement spectrum as noted above. 

\vspace{5mm}

\section{Topological Entanglement Entropy Computation for general $t, \mu, \Delta$}
\label{app:TEEexact}
In this appendix, we discuss results for the topological entanglement entropy obtained for projected wave functions at arbitrary $t, \mu, \Delta$ beyond the $t = \Delta$ and $\mu = 0$ limit considered in the main text; the results are shown in Fig.~\ref{fig:topEEfig}. Using exact diagonalization methods, we compute the TEE of the mean-field wave function computed at arbitrary $t, \mu, \Delta$ that is then projected to a fixed particle number represented by $p$, the filling factor, in the plots. In Fig. \ref{fig:topEEfig} of the main text, we showed that when fixing $\mu = 0$ one finds good agreement with the theoretical prediction in Eq. \ref{eq:topEEcalc} at filling fractions away from $p = 0, 1,$ barring finite size deviations. Varying $\mu$, we find that away from the transition, the topological entanglement entropy again robustly clings on to the theoretical prediction from $\mu = 0$ for a large range of filling fractions $p$ away from $p = 0,1$. Beyond the transition into the trivial regime at $\abs{\mu} > 2t$, the entanglement entropy indeed starts deviating significantly from the theoretical prediction in the topological regime. 

Note that in general, away from the special $t = \Delta, \mu = 0$ point, the ground state will be only approximately doubly degenerate in the topological phase, while the gap is finite in the trivial phase. Since all eigenstates have definite parity, we note that we can only extract wave functions with fixed particle number $\ket{N}$ (and associated filling fractions $p$) if they have the same parity as the true ground state. However, in the plots below, we show the results for the TEE even for wave functions $\ket{N}$ that correspond to the opposite parity as the true ground state. In the topological regime, since the degeneracy is exponentially small in the system size, the TEE $\mathcal{S}_{\text{topo}}$ is approximately the same for wave functions $\ket{N}$ with opposite parity. In the non-topological regime, however, we see significant oscillations in  $\mathcal{S}_{\text{topo}}$ as the gap becomes finite. The lower values in the oscillations correspond to the states $\ket{N}$ extracted from the true ground state. 


Further examining the plots, we note that if we keep $\mu = 0$ fixed and vary $t \neq \Delta$, we find that for a large range of fillings, the saturation value of the TEE deviates from the value for $t = \Delta$. However, this is a finite-size effect, as can be observed by comparing the computation of the TEE for $L = 16$ and $L = 20$. Finally, in order to probe the non-universal piece of the TEE, we plot $\mathcal{S}_{\text{topo}}$ for various entanglement partitions $(L_{A}, L_{B}, L_{C}, L_{D});$ the numerics show very good agreement with Eq. \ref{eq:topEEcalc}.

\begin{widetext}
    \begin{figure}
        \includegraphics[width=\textwidth]{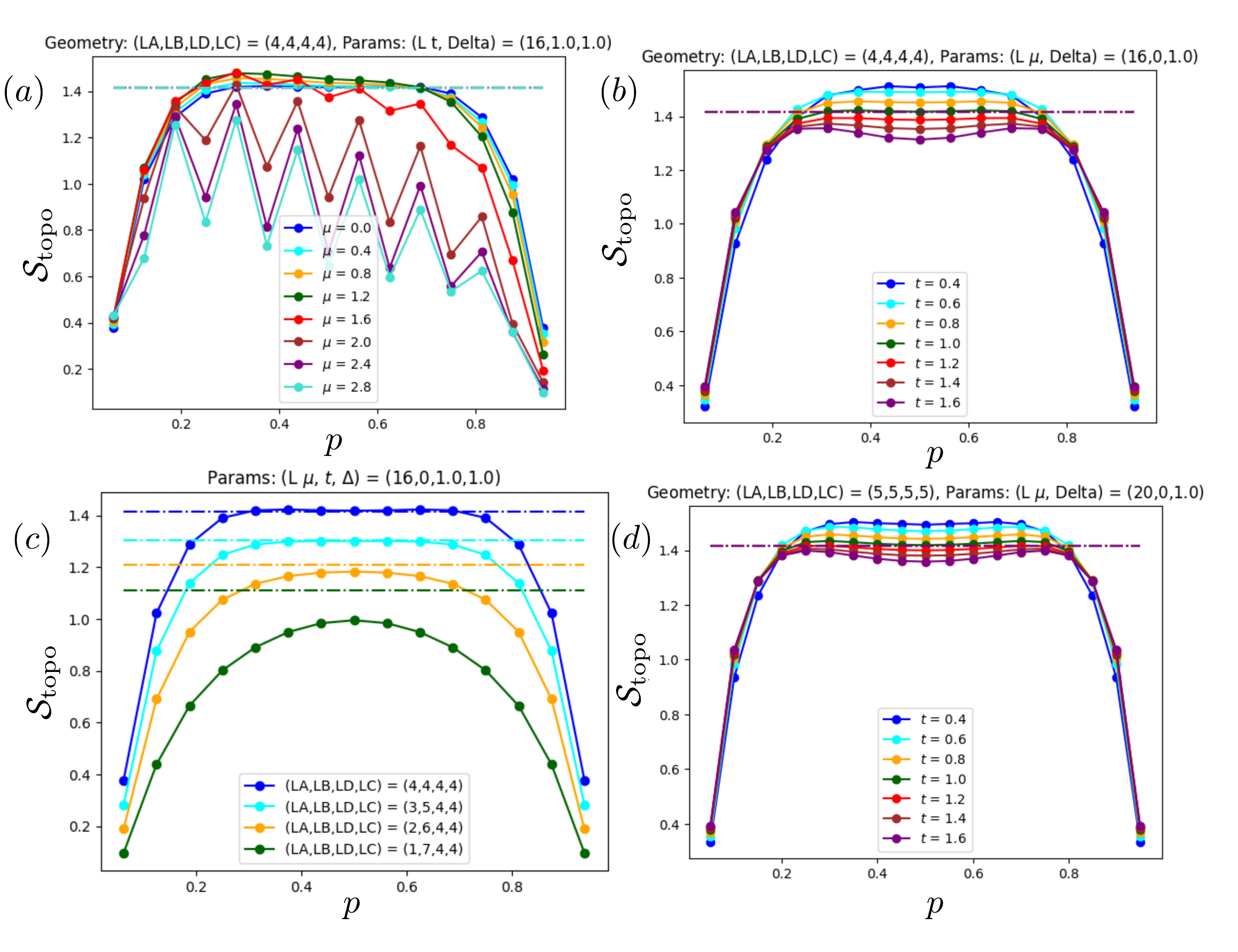}
        \caption{Calculation of the topological entanglement entropy using Exact diagonalization for (a) fixed $t = \Delta = 1, L = 16$, varying $\mu$; (b) and (d) fixed $\mu = 0, \Delta = 1$, $L = 16$ and $L = 20$ respectively; and (c) fixed $\mu = 0, t = \Delta = 1$ and different partitions. Dashed lines correspond to the prediction according to Eq.~(\ref{eq:topEEcalc}) of the main text.}
        \label{fig:topEEfig}
    \end{figure}
\end{widetext}

\end{document}